%% file: main.tex
\documentclass[sigplan,10pt,nonacm]{acmart}

\renewcommand\footnotetextcopyrightpermission[1]{}
\AtBeginDocument{  }

\usepackage[dvipsnames]{xcolor}
\usepackage{threeparttable}
\usepackage{tabularx}
\usepackage{makecell}
\usepackage{booktabs}
\usepackage[most]{tcolorbox}
\usepackage{enumitem}
\usepackage{fancyhdr}

\newtcolorbox{tightbox}{
    colframe=black,
    width=\linewidth,
    arc=2mm,
    auto outer arc,
    breakable,
    before skip=4pt,
    after skip=4pt,
    top=1pt,          bottom=1pt,       left=3pt,
    right=3pt,
        }

\newcommand{\prjname}{\textsf{CHASE}}

\newcommand{\authorfootnotes}{  \begingroup
  \renewcommand{\thefootnote}{*}  \footnotetext{These authors contributed equally to this work.}  \endgroup
  \begingroup
  \renewcommand{\thefootnote}{**}  \footnotetext{These authors are corresponding authors.}  \endgroup
}
\begin{document}

\title{\texorpdfstring{Application-Driven Architecture Exploration for Cross-Layer Heterogeneous Systems}{CHASE: Application-Driven Architecture Exploration for Cross-Layer Heterogeneous Systems}}

\author{Yuchen Fan}
\author{Minghong Sun}
\author{Jikui Ma}
\author{Yunpeng Xu}
\author{Shunyu Mao}
\author{Liu He}
\author{Shunan Dong}
\author{Jiahao Yang}
\author{Yu Zhu}
\author{Xinhao Yang}
\author{Tianyan Zhong}
\author{Haoran Sun}
\author{Daoqi Liu}
\author{Zongle Huang}
\author{Xinyuan Lin}
\author{Huazhong Yang}
\author{Maokun Li}
\author{Yongpan Liu}
\author{Yu Wang}
\affiliation{  \institution{Tsinghua University}
}

\author{Zhenhua Zhu}
\affiliation{  \institution{Tsinghua University}
}

\author{Hongyang Jia}
\author{Shuwen Deng}
\affiliation{  \institution{Tsinghua University}
}

\renewcommand{\shortauthors}{Fan et al.}

\makeatletter
\def\@mkauthors{  \global\setbox\mktitle@bx=\vbox{    \noindent\unvbox\mktitle@bx
    \vskip 0.8em
    \centering
    {\@authorfont
      Yuchen Fan\textsuperscript{1,*}\quad
      Minghong Sun\textsuperscript{1,*}\quad
      Jikui Ma\textsuperscript{1,*}\quad
      Yunpeng Xu\textsuperscript{1,*}\quad
      Shunyu Mao\textsuperscript{1,*}\par
      Liu He\textsuperscript{1}\quad
      Shunan Dong\textsuperscript{1}\quad
      Jiahao Yang\textsuperscript{1}\quad
      Yu Zhu\textsuperscript{1}\quad
      Xinhao Yang\textsuperscript{1}\par
      Tianyan Zhong\textsuperscript{1}\quad
      Haoran Sun\textsuperscript{1}\quad
      Daoqi Liu\textsuperscript{1}\quad
      Zongle Huang\textsuperscript{1}\quad
      Xinyuan Lin\textsuperscript{1}\par
      Huazhong Yang\textsuperscript{1}\quad
      Maokun Li\textsuperscript{1}\quad
      Yongpan Liu\textsuperscript{1}\quad
      Yu Wang\textsuperscript{1}\par
      Zhenhua Zhu\textsuperscript{1,**}\quad
      Hongyang Jia\textsuperscript{1,**}\quad
      Shuwen Deng\textsuperscript{1,**}\par
    }    \vskip 0.6em
    {\@affiliationfont
      \textsuperscript{1}Tsinghua University\par
    }    \vskip 0.8em
  }}
\makeatother

\input{sec/00_abstract}


\settopmatter{printacmref=false} 
\renewcommand\footnotetextcopyrightpermission[1]{} 
\maketitle
\authorfootnotes
\pagestyle{plain}

\input{sec/01_introduction}

\input{sec/02_bgarw}

\input{sec/03_overview}

\input{sec/04_modeling}
\input{sec/05_cod}
\input{sec/06_calib}
\input{sec/07_eval}

\input{sec/08_con}

\clearpage

\bibliographystyle{ACM-Reference-Format}
\bibliography{references}

\appendix
\input{sec/10_app}
\input{sec/075_rnd}

\end{document}

%% file: sec/00_abstract.tex
\begin{abstract}
AI and HPC infrastructure increasingly serves workload portfolios that combine dense tensor computation, sparse kernels, large memory footprints, and communication-intensive collectives. Supporting such portfolios requires coordinated choices across accelerators, memory tiers, scale-up fabrics, and cluster networks. The resulting Cross-layer Heterogeneous System (XHS) design space is difficult to explore: hardware choices change the legal task mappings, while rack power, switch radix, cabling, and cost constraints invalidate many candidates. Existing node-scale design-space exploration tools and fixed-platform distributed simulators address only parts of this problem.

We present \prjname{}, an application-driven framework that searches physically feasible XHS architectures through the workloads they must execute. \prjname{} represents candidates as hierarchical typed graphs and rejects designs that violate deployment constraints. It avoids an intractable joint hardware-mapping search with a decoupled two-level loop: an inner mapper translates hardware-independent workload DAGs into topology-aware event traces, a calibrated event-driven simulator evaluates each mapping, and an outer telemetry-guided optimizer evolves the hardware graph.

We evaluate \prjname{} on sparse-computing and LLM workloads. Its mapper remains within 6.06\% of exhaustive optima while reducing mapping time by 60.5\% on average relative to PEFT. Compute-model errors average 4.4--7.5\%, and communication validation reproduces key trends across physical platforms. The outer search reaches near-global optima within 64 iterations. End-to-end case studies show that sparse workloads favor criticality-aware heterogeneous pods, whereas LLM inference favors scale-up islands; the resulting designs deliver 6.20$\times$ and 2.12$\times$ geomean speedups, respectively, while reducing cost and power relative to the baselines.
\end{abstract}

%% file: sec/01_introduction.tex
\section{Introduction}

AI and high-performance computing (HPC) infrastructure increasingly serves portfolios rather than a single workload class. Trillion-parameter large language models (LLMs) and extreme-scale simulations can require thousands of PFLOPs and tens of terabytes of memory, pushing execution from individual chips and servers to large clusters~\cite{sevilla2022compute, reed2023hpc, narayanan2021megatron, williams2009spmv}. Their resource profiles, however, differ substantially across compute, memory, and communication.

At the same time, world models~\cite{nvidia2025ces} and HPC-AI convergence~\cite{huerta2020convergence} place these different phases in shared, closed-loop pipelines. Emerging \textit{AI Factories} combine simulation, data processing, training, and inference instead of operating them as isolated services~\cite{nvidia2026aifactories}. Autonomous-agent development, for example, cycles through synthetic-data rendering, vision-language-action training, and real-time inference~\cite{brohan2023rt2, nvidia2025ces}. AI-assisted scientific discovery similarly connects
sparse solvers and large-scale simulations that generate physical-state data, AI models that choose the next simulation batch, and inference services that screen candidates online~\cite{wang2023scientificdiscovery, tran2018activelearning}.

Serving these mixed pipelines motivates \textbf{Cross-layer Heterogeneous Systems (XHS)} that combine CPUs, GPUs, and domain-specific accelerators; multi-tier memory such as UnifiedBus~\cite{ub2025} and CXL~\cite{cxl31}; and high-radix networks spanning package, node, rack, and cluster scales~\cite{jouppi2023tpuv4, li2023pond}. In this paper, ``\textit{cross-layer}'' denotes joint reasoning over workload behavior, node-level compute and memory composition, and pod- or cluster-level interconnect and deployment constraints. Our question is architectural: \textit{how should compute devices, memory tiers, and networks be composed and connected for a target workload portfolio?}

We present \textbf{\prjname{}}, an application-driven architecture exploration framework for XHS. Given target workloads, a hardware design space, and physical constraints, \prjname{} returns an optimized, deployable architecture ranked using candidate-specific workload mappings and event-level performance telemetry. Two coupled obstacles shape this exploration problem.

\textbf{Challenge 1 (C1): Different domains of workflows contain diverse phases that stress different system resources.}
No single resource mix is optimal across the workload portfolio. Our cross-workload evaluation makes the mismatch concrete: the LLM-optimized architecture is $1.91\times$ slower on HPC workloads than the HPC-optimized design, whereas the HPC-optimized architecture is $2.77\times$ slower on LLM workloads than the LLM-optimized design.
LLM pre-training emphasizes dense tensor throughput and collective bandwidth~\cite{narayanan2021megatron, zheng2022alpa}, whereas sparse-matrix and graph computations are often limited by irregular memory accesses, reductions, and low-latency data movement~\cite{abadal2021computing, ballard2014communication, williams2009spmv}. A fixed, one-size-fits-all architecture consequently strands different resources in different phases.
Converged deployments therefore require a system-level compromise that serves heterogeneous phases under shared budget, power, and operational constraints.

\textbf{Challenge 2 (C2): physically constrained XHS design.} The candidate space is massive and discrete: systems vary in device ratios, scale-up fabrics, memory-pooling degrees, and scale-out topologies. These choices are coupled to deployment constraints. Adding accelerators may increase peak throughput, but it also raises rack power, changes communication paths, and can violate cabling or switch-radix limits. Industry systems are therefore commonly assembled from vendor-defined units such as NVIDIA DGX~\cite{nvidia2023dgxsuperpod}, AMD MI350~\cite{amd2025mi350}, and TPU v4 Pods~\cite{jouppi2023tpuv4}. Although effective for their intended workloads, these fixed point solutions can overprovision compute for sparse or converged workloads and reduce cost efficiency~\cite{abadal2021computing, williams2009spmv}.

Existing tools cover only one side of this problem. Accelerator DSE frameworks such as MAGNet and Timeloop~\cite{kwon2019maestro, parashar2019timeloop} search bounded node-scale spaces without cluster-level constraints such as switch radix and rack power. Distributed-training simulators such as ASTRA-SIM~\cite{rashidi2020astrasim, won2023astrasim2} provide scalable evaluation but assume a fixed platform. This separation exposes a \textit{topology-mapping deadlock}: a hardware point $h$ determines the resources, contention domains, and routes that define its legal mapping space $S_h$. Hardware cannot be ranked fairly without an effective mapping, yet the mapping cannot be constructed before the hardware is known. A flat joint search over both spaces is intractable. \prjname{} breaks this cycle with two mechanisms.

\noindent\textbf{Solution 1 (S1): Decoupled Two-Level Optimization.} \prjname{} places candidate-specific mapping in an inner loop and hardware evolution in an outer loop. For each architecture, the \textit{Mapper} lowers hardware-independent workload DAGs into topology-aware event traces, and an event-driven \textit{Simulator} measures their performance. The TG-RL-driven \textit{Optimizer} then uses normalized bottleneck telemetry to propose the next feasible hardware candidate. This decomposition preserves hardware-mapping dependence without forming one monolithic search space.

\noindent\textbf{Solution 2 (S2): Hierarchical Graph Modeling and Constraint Filtering.} \prjname{} represents hardware as typed graphs organized across package, node, rack, and cluster levels. Structured device and interconnect templates replace a flat enumeration of components. Predicates over budget, power, rack space, cable length, and switch radix reject invalid graphs \textit{before} mapping and simulation, so the optimizer spends its evaluation budget only on deployable candidates while retaining the explicit relationship between $h$ and $S_h$.

Because architecture rankings are only as reliable as their timing model, \prjname{} also includes a measurement-driven calibration path. It fits event parameters to commercial-platform measurements spanning NVLink, PCIe, and InfiniBand, then checks scale-out behavior across device counts, message sizes, and communication fanouts. These measurements ground the simulator before it is used inside the exploration loop.

We evaluate both the individual components and the end-to-end search. On tractable mapping instances, \prjname{} stays within 6.06\% of exhaustive optima while reducing mapping time by 60.5\% on average relative to PEFT. Compute-model errors average 4.4--7.5\%, intra-machine communication errors average below 10\%, and a held-out platform has 5.8\% mean error. The optimizer reaches near-global architectures within 64 iterations. Finally, the discovered sparse-computing and LLM systems achieve 6.20$\times$ and 2.12$\times$ geomean speedups, respectively, while lowering cost and power relative to the baselines.

This paper makes the following contributions:
\begin{itemize}[leftmargin=*,topsep=3pt,itemsep=2pt,parsep=0pt,partopsep=0pt]
    \item \textbf{A constrained XHS formulation.}
    We formulate architecture exploration as a search over hierarchical hardware graphs whose legal workload-mapping spaces depend on topology. Explicit physical constraints remove undeployable candidates before expensive evaluation.
    \item \textbf{An end-to-end exploration framework.} To the best of our knowledge, \prjname{} is the first end-to-end DSE framework at XHS scale. Its decoupled two-level optimization integrates a workload \textit{Mapper}, an event-driven \textit{Simulator}, and a TG-RL-driven \textit{Optimizer} to resolve the topology-mapping deadlock.
        \item \textbf{Calibration.}
    We fit event models to commercial-cluster measurements and validate both point-level timing and scale-out trends, grounding architecture rankings in physical data.
    \item \textbf{Component and end-to-end evaluation.}
    We quantify mapping quality, simulation fidelity, and optimizer convergence, then use sparse-computing and LLM case studies to expose distinct workload-dependent XHS designs with 6.20$\times$ and 2.12$\times$ geomean speedups.
\end{itemize}

%% file: sec/02_bgarw.tex
\section{Background, Related Work and Motivation}
\label{sec:background_motivation}

The necessity for a systematic and agile architectural exploration framework is driven by an inescapable conflict in the current computing landscape: the extreme diversity of emerging workload demands against the massive and highly constrained design space of modern hardware.

\begin{figure}[t]
    \centering
    \includegraphics[width=0.8\linewidth]{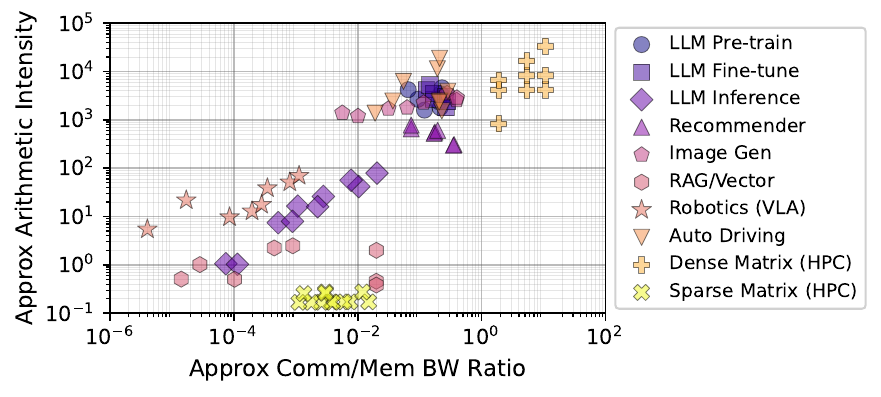}
    \caption{Characteristics of diverse HPC and AI workloads, mapped by Arithmetic Intensity (FLOPs/Byte) and Communication-to-Memory Bandwidth Ratio. The variance strictly demonstrates that a hardware configuration optimal for one workload paradigm may perform poorly on another. 
        }
    \label{fig:workload_characteristics}
\end{figure}

\subsection{The Diversified Demands of Emerging Workloads (C1)}
As artificial intelligence and high-performance computing converge into closed-loop workflows, computational tasks no longer exhibit uniform execution patterns. Instead, different phases of these workflows stress vastly different system resources, resulting in profound workload diversity (\textbf{Challenge 1 (C1)}).

As illustrated in Figure~\ref{fig:workload_characteristics}, distinct computational paradigms map to completely different regions of the operational space. 
        
    \textbf{Dense LLM Workloads:} Tasks such as large language model inference typically present high arithmetic intensity. They relentlessly stress dense tensor throughput and require massive collective communication bandwidth. \cite{narayanan2021megatron, zheng2022alpa}.
        
    \textbf{Sparse Matrix Computation:} Representing a vital class of scientific and HPC applications, sparse matrix computation (e.g., iterative solvers represented by the HPCG benchmark~\cite{doi:10.1177/1094342015593158}) is memory-bound and control-intensive. These workloads exhibit low arithmetic intensity and are fundamentally bottlenecked by operations like sparse matrix-vector multiplication (SpMV), irregular memory access patterns, and latency-sensitive fine-grained point-to-point (P2P) data movement. Consequently, they require extreme memory bandwidth and low-latency scalar processing rather than peak floating-point throughput \cite{ williams2009spmv}.

Because the compute-to-communication ratio and data-movement behaviors vary so drastically, a fixed hardware architecture tuned for one workload regime suffers from severe resource underutilization or critical network bottlenecks when executing another. This stark diversity invalidates the traditional ``one-size-fits-all'' hardware approach.

\begin{figure}[t]
    \centering
    \includegraphics[width=0.8\linewidth]{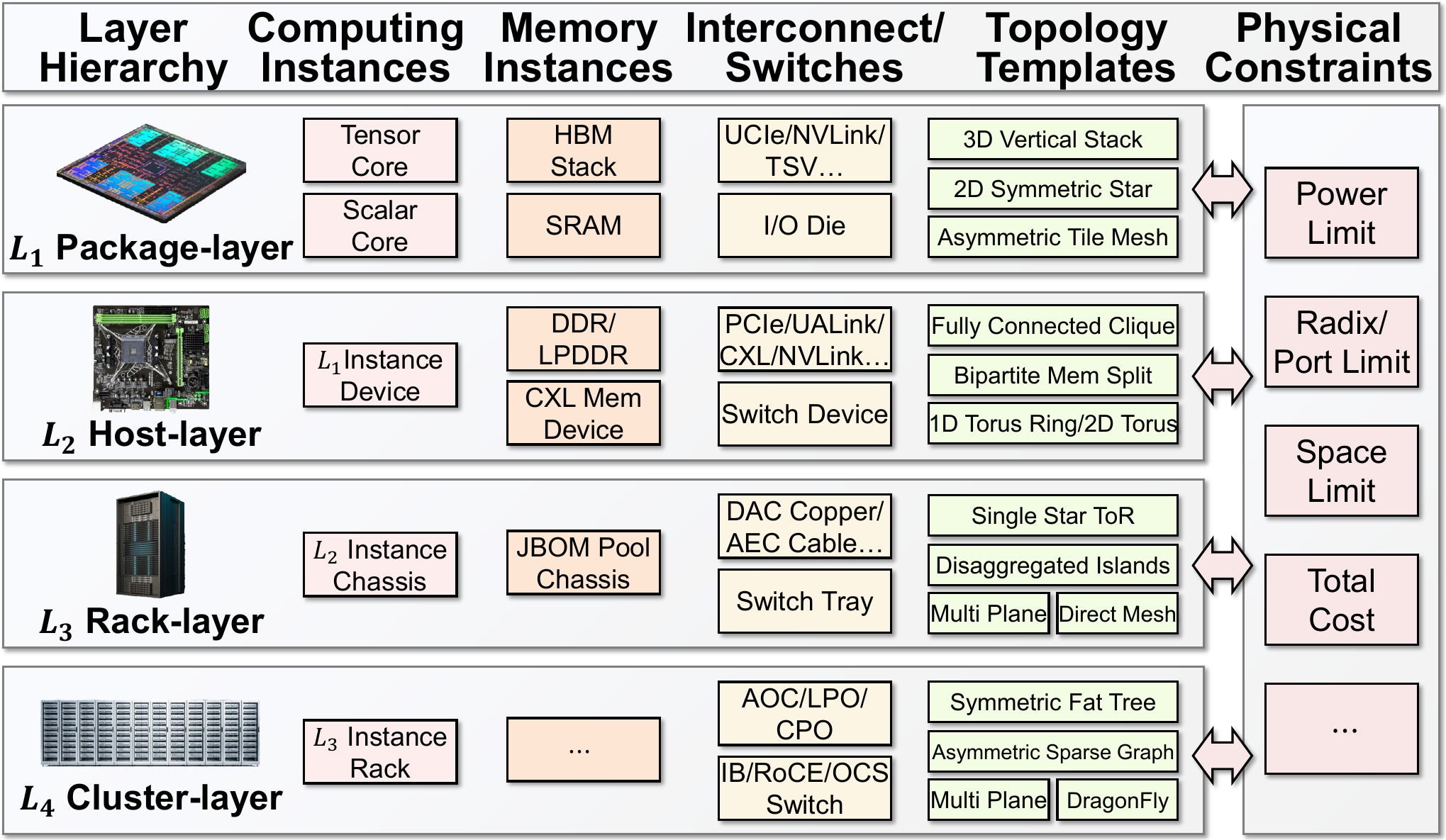}
        \caption{The 4-layer hierarchical architecture of a typical XHS (Package, Node, Rack, Cluster). Physical constraints such as power, thermal limits, and wiring feasibility strictly bound the design space at each layer.}
        \label{fig:xhs_architecture}
\end{figure}

\begin{table}[t]
\centering
\begin{threeparttable}
\caption{
Commercial XHS instantiations exhibit diverse choices in organization,
interconnect, topology, and per-unit capability.
Notation:
(1) {Cluster} denotes compute racks : communication racks;
(2) {Rack} denotes compute nodes : switch trays, except AL128 where it denotes
accelerators : CPU nodes;
(3) {Node} denotes accelerators : CPUs;
(4) {Profile} reports FP16 PFLOPS / I/O BW / HBM BW / HBM capacity per accelerator unit;
(5) {Topo.}: 1L/2L = 1-/2-layer, MP = multi-plane,
Orth. = orthogonal, FC = fully connected.
}
\label{tab:commercial_superpods}
\scriptsize
\setlength{\tabcolsep}{2.2pt}
\renewcommand{\arraystretch}{1.12}
\begin{tabularx}{\columnwidth}{@{}l c c c l c X@{}}
\toprule
\textbf{System}
&
\textbf{Cluster}
&
\textbf{Rack}
&
\textbf{Node}
&
\textbf{Fabric}
&
\textbf{Topo.}
&
\textbf{Profile per Acc.}
\\
\midrule
Huawei CM384
&
12:4
&
4:1
&
8:4
&
UB
&
2L-MP
&
0.8 / 0.78 / 1.6 / 128
\\
Alibaba AL128
&
--
&
32:16\tnote{1}
&
4:--\tnote{2}
&
ALink
&
1L-Orth.
&
--\tnote{2} / 0.8 / --\tnote{2} / 144
\\
NVIDIA GB200
&
--
&
18:9
&
4:2
&
NVLink5
&
1L-FC
&
5 / 1.8 / 8 / 192
\\
AMD MI350
&
--
&
8$\sim$16:0$\sim$4
&
8:1
&
IF
&
1L-FC
&
4.6 / 1.08 / 8 / 288
\\
AMD Helios
&
--
&
18:6
&
4:1
&
UALoE
&
1L-FC
&
10 / 4.8 / 19.6 / 432
\\
\bottomrule
\end{tabularx}

\begin{tablenotes}
\item[1] For AL128, the rack-level ratio denotes accelerators : CPU nodes. The CPU count per accelerator node is not publicly specified.
\item[2] Not publicly available or unspecified.
\end{tablenotes}
\end{threeparttable}
\end{table}

\begin{table*}[t]
\centering
\begin{threeparttable}
\caption{
Comparison of representative DSE and system co-design frameworks.
\(\checkmark\): core support; \(\circ\): partial or parameterized support; --: out of scope.
``Phys. cons.'' denotes the physical feasibility boundary. 
}
\label{tab:dse_comparison_compact}
\scriptsize
\setlength{\tabcolsep}{3.2pt}
\renewcommand{\arraystretch}{1.12}
\begin{tabularx}{\textwidth}{@{}l l c c c c l l l@{}}
\toprule
\textbf{Framework}
&
\textbf{Target}
&
\textbf{Level}
&
\textbf{HW-DSE}
&
\textbf{Map-DSE}
&
\textbf{Topo-DSE}
&
\textbf{Phys. cons.}
&
\textbf{Eval.}
&
\textbf{Workloads}
\\
\midrule
MAGNet~\cite{venkatesan2019magnet}
&
NN accel.
&
ASIC
&
\(\checkmark\)
&
\(\checkmark\)
&
--
&
PPA
&
RTL/model
&
DNN
\\
MAESTRO~\cite{kwon2019maestro}
&
Dataflow
&
ASIC
&
\(\circ\)
&
\(\checkmark\)
&
--
&
PPA
&
Analytic
&
DNN
\\
Timeloop / Accelergy~\cite{parashar2019timeloop,wu2019accelergy}
&
Tensor accel.
&
ASIC/node
&
\(\checkmark\)
&
\(\checkmark\)
&
--
&
PPA
&
Analytic
&
Tensor
\\
Sparseloop~\cite{wu2022sparseloop}
&
Sparse accel.
&
ASIC/node
&
\(\checkmark\)
&
\(\checkmark\)
&
--
&
PPA
&
Analytic
&
Sparse
\\
GAMMA / ConfuciuX / DOSA~\cite{kao2020gamma,kao2020confuciux,hong2023dosa}
&
DNN optimizer
&
ASIC
&
\(\checkmark\)
&
\(\checkmark\)
&
--
&
PPA
&
Cost model
&
DNN
\\
ASTRA-sim2.0~\cite{won2023astrasim2}
&
Training sim.
&
Cluster
&
\(\circ\)
&
\(\circ\)
&
\(\circ\)
&
--
&
Trace sim.
&
DL/LLM
\\
TopoOpt/LIBRA~\cite{wang2023topoopt,won2024libra}
&
Network DSE
&
Cluster
&
\(\circ\)
&
\(\checkmark\)
&
\(\checkmark\)
&
Network
&
Sim/proto
&
DNN train
\\
vTrain / Calculon / AMPeD~\cite{bang2024vtrain,isaev2023calculon,moolchandani2023amped}
&
LLM config.
&
Cluster
&
\(\circ\)
&
\(\circ\)
&
\(\circ\)
&
Cost
&
Profile/model
&
LLM
\\
\textbf{\prjname{}}
&
\textbf{XHS DSE}
&
\textbf{Pkg--cluster}
&
\(\checkmark\)
&
\(\checkmark\) \(\boldsymbol{S_h}\)\tnote{1}
&
\(\checkmark\)
&
\textbf{Pwr/radix/cable}
&
\textbf{Point+trend}
&
\textbf{Sparse+LLM}
\\
\bottomrule
\end{tabularx}

\begin{tablenotes}
\item[1] Unlike prior accelerator-level DSE tools or cluster-level training simulators, \prjname{} searches physically feasible XHS hardware graphs and evaluates each candidate only after constructing its topology-dependent mapping space \(S_h\).
\end{tablenotes}
\end{threeparttable}
\end{table*}

\subsection{The Rise and Fragmentation of Cross-Layer Systems (C2)}
To sustain the diverse demands in \textbf{C1}, the computing infrastructure has evolved far beyond scale-up single nodes into \textbf{Cross-layer Heterogeneous Systems (XHS)}. As shown in Figure~\ref{fig:xhs_architecture},
a typical XHS is an orchestrated hierarchy spanning four critical physical boundaries:
1) \textit{Package layer}: heterogeneous logic dies and co-packaged interconnects;
2) \textit{Node layer}: multi-tier memory and PCIe/CXL fabrics;
3) \textit{Rack layer}: disaggregated resource pooling and intra-rack switches constrained by power delivery;
4) \textit{Cluster layer}: high-radix networks (e.g., InfiniBand, Ethernet) and optical transceivers.

Currently, the industry attempts to tackle this massive design space through point-designed commercial products, often marketed as ``Pods'' or ``SuperPODs.'' However, instead of converging on a standardized architecture, state-of-the-art XHS instantiations from major vendors exhibit significant fragmentation (Table~\ref{tab:commercial_superpods}). At the node layer, the ratio of accelerators to host CPUs varies widely (e.g., 8:4 in Huawei CloudMatrix vs. 8:1 in AMD MI350 Series \cite{amd2025mi350}). Furthermore, cluster-level routing topologies range from single-layer fully-connected cliques to multi-plane orthogonal networks.

This fragmentation shows that the XHS design space is highly non-linear and constrained (\textbf{C2}). The coupling of discrete topological choices with strict physical limitations (e.g., rack power caps and cable reach) creates a massive search space where manual tuning and trial-and-error prototyping easily settle in suboptimal local minima.

\subsection{The Topology-Mapping Deadlock in Current DSE}
While application-driven XHS customization is urgently needed to resolve the conflict between \textbf{C1} and \textbf{C2}, existing Design Space Exploration (DSE) frameworks fall short of supporting this cross-layer scenario.

Table~\ref{tab:dse_comparison_compact} shows some presentative DSE and system co-design frameworks. Traditional accelerator DSE tools (e.g., MAGNet, Timeloop, MAESTRO \cite{kwon2019maestro, parashar2019timeloop, venkatesan2019magnet}) focus heavily on bounded ASIC- or node-scale parameterization. They lack the abstraction for cluster-scale organization and ignore critical physical constraints such as switch radix, cable lengths, and rack power envelopes. Conversely, distributed-system simulators (e.g., ASTRA-SIM \cite{rashidi2020astrasim, won2023astrasim2}) provide scalable evaluation for extreme-scale runtimes, but they inherently require a fixed, pre-defined hardware platform and a pre-compiled execution trace to function.

This disparity exposes a fundamental hurdle that prevents existing tools from solving \textbf{C1} and \textbf{C2}, which we identify as the topology-mapping deadlock:
Hardware topology fundamentally determines the software mapping space. We cannot accurately evaluate a hardware candidate without a valid workload mapping, yet we cannot generate a valid mapping (tensor partitioning, collective scheduling, and routing) without knowing the specific devices, links, and contention domains exposed by that hardware candidate.

Evaluating software mappings on a single fixed platform fails to compare alternative hardware architectures fairly. Therefore, to break this deadlock 
and effectively co-optimize across the workload (\textbf{C1}) and system (\textbf{C2}) spaces, an agile framework that \textit{simultaneously} evolves physically feasible hardware graphs and explores their corresponding topology-dependent software mappings is urgently required.

%% file: sec/03_overview.tex
\section{Framework Overview}
\label{sec:framework_overview}

\begin{figure}[t]
    \centering
    \includegraphics[width=0.8\linewidth]{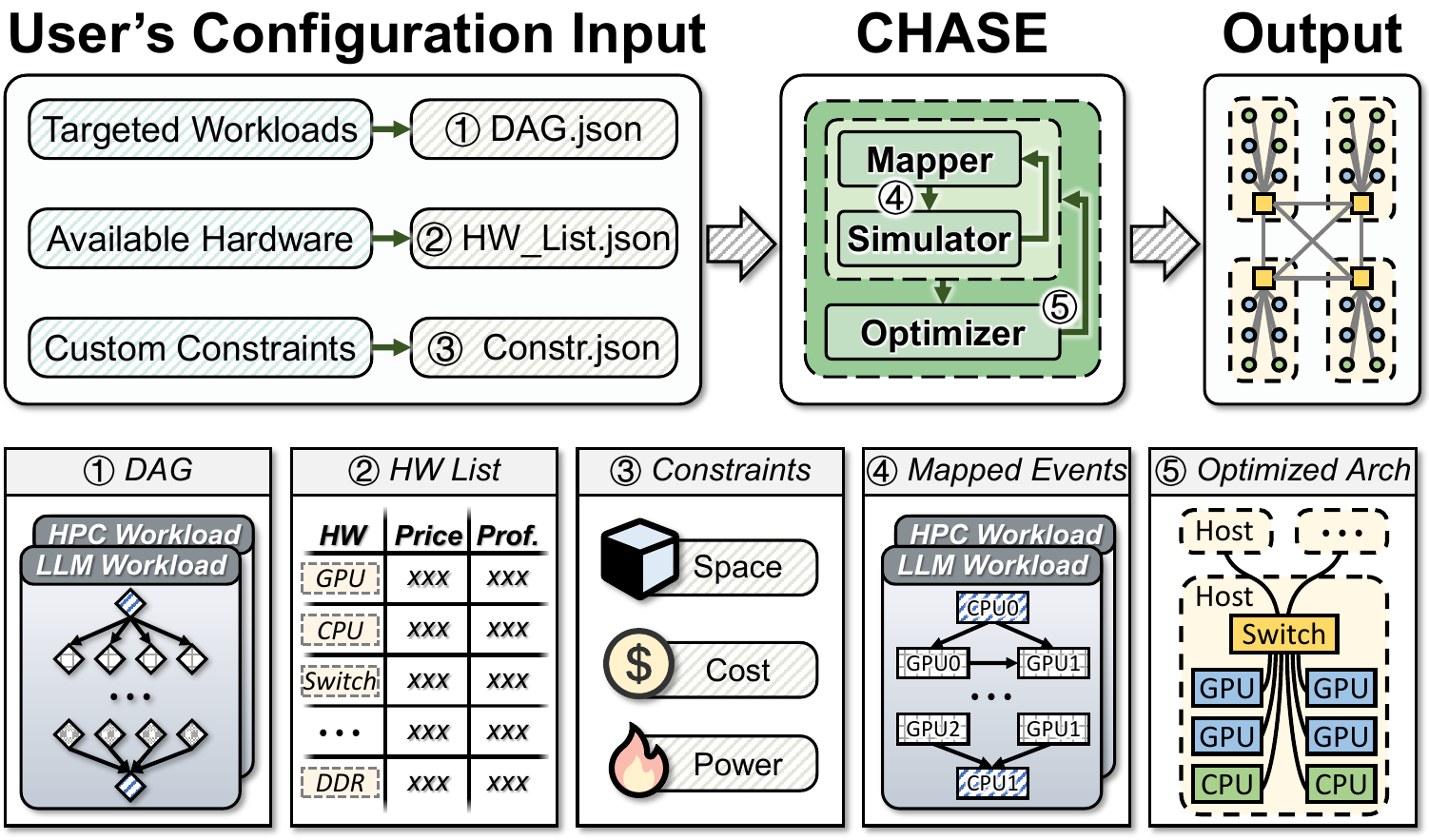}
    \caption{User-facing input/output contract of \prjname{}. Users provide target workload DAGs, an available-hardware catalog, and deployment constraints; \prjname{} maps the workloads onto candidate hardware, evaluates them through simulation, and returns an optimized XHS architecture.}
    \label{fig:instruction}
\end{figure}

\begin{figure*}[t]
    \centering
    \includegraphics[width=0.8\textwidth]{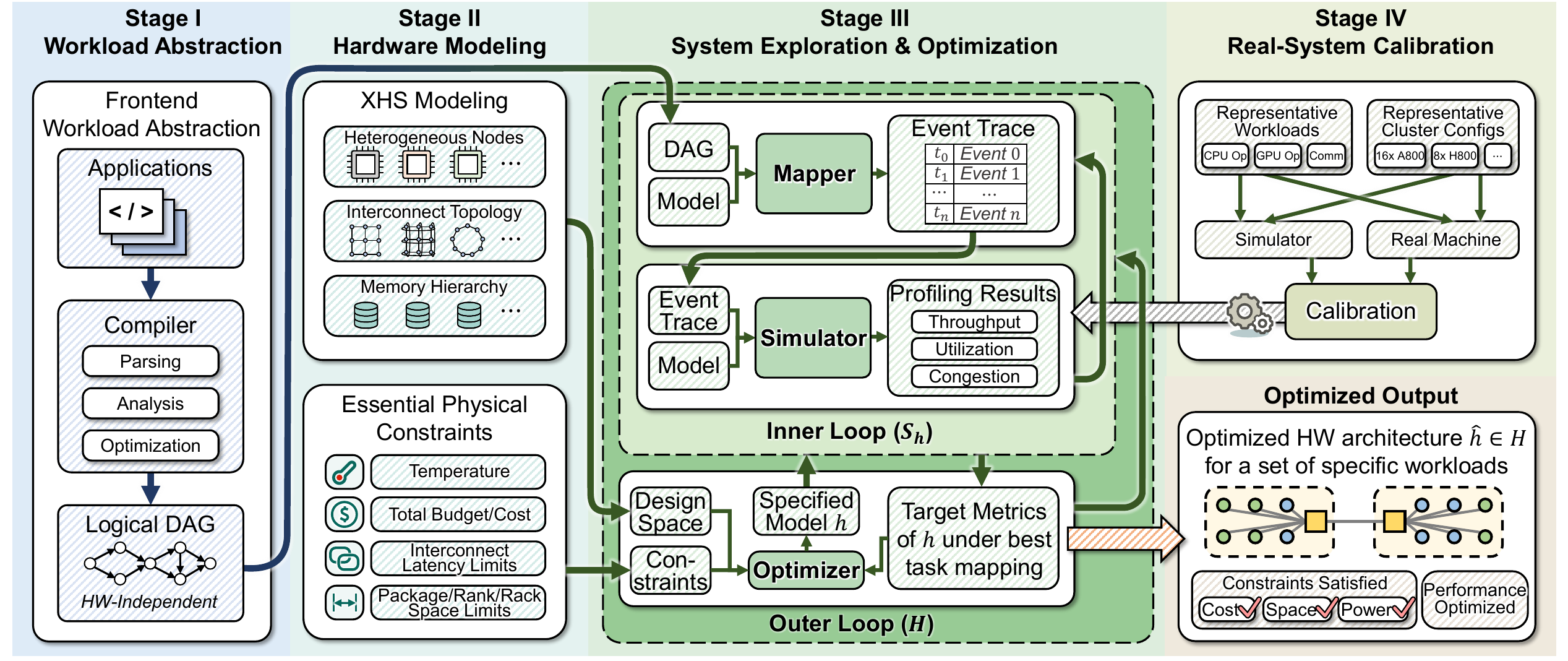}
    \caption{{The end-to-end architecture of \prjname{}. The framework abstracts workloads into hardware-independent DAGs (Stage I), models physically constrained hardware graphs (Stage II), employs a decoupled two-level optimization loop (Stage III) to iteratively search topologies and software mappings, guided by a hardware-validated event-driven simulator (Stage IV).}}
    \label{fig:framework_overview}
\end{figure*}

We propose \textbf{\prjname{}}, an application-driven architecture exploration framework for cross-layer heterogeneous systems. 
Figure~\ref{fig:instruction} summarizes its user-facing contract: users specify target workloads, hardware candidates, and constraints, while \prjname{} returns topology-aware mapped events and an optimized architecture.
As illustrated in Figure~\ref{fig:framework_overview}, the framework bridges workload characteristics and physical constraints via a decoupled two-level optimization pipeline grounded in real-system calibration.

\prjname{} addresses the challenges identified in Section~\ref{sec:background_motivation}. For the physically constrained XHS space (\textbf{C2}), it models architectures as hierarchical typed graphs and filters them with physical feasibility constraints. For diverse workloads (\textbf{C1}), it abstracts applications into hardware-independent DAGs and instantiates them onto candidate-specific topologies. 

\prjname{} employs three technical mechanisms.
First, each hardware candidate $h$ defines its own topology-dependent mapping space $S_h$, preventing invalid single-platform comparisons.
Second, it applies physical constraint predicates to hierarchical hardware graphs before mapping and simulation.
Third, it uses simulator telemetry to guide the outer-loop optimizer toward architecture edits that relieve measured workload bottlenecks.

\subsection{End-to-End Exploration Workflow}
To align with the architectural components in Figure~\ref{fig:framework_overview}, the \prjname{} workflow is organized into four interconnected stages:

\textbf{Stage I: Hardware-Independent Workload Abstraction.} To reduce hardware-specific noise during the initial phase, \prjname{} employs a compiler frontend to parse and analyze the target applications (e.g., HPC sparse solvers, LLMs). The application is abstracted into a hardware-neutral Directed Acyclic Graph (DAG)~\cite{topcuoglu2002heft,sridharan2026mlcommonschakra}. This logical DAG retains algorithmic behaviors, such as computation volume (FLOPs) and data dependency sizes, keeping the workload representation independent of any specific hardware. 

\textbf{Stage II: Hardware Modeling and Constraint Filtering.} \prjname{} treats  XHS design space as a hierarchy of typed hardware graphs spanning package, node, rack, and cluster levels. Crucially, before any expensive simulation occurs, an essential \textit{Physical Constraint Verifier} filters out invalid hardware configurations that violate power envelopes, total budget, switch radix boundaries, or volumetric space limits.

\textbf{Stage III: System Exploration \& Optimization.} The core of \prjname{} lies in its exploration engine. A dedicated \textit{Mapper} projects the hardware-independent DAG onto a physically feasible hardware topology candidate, generating a hardware-aware event trace. Subsequently, a \textit{Hardware-Calibrated Simulator} executes this trace to evaluate resource contention, queueing, and network bottlenecks~\cite{casanova2014simgrid,wilke2015sstmacro,rashidi2020astrasim}. The \textit{Optimizer} then uses this telemetry to dynamically propose evolved hardware candidates.

\textbf{Stage IV: Real-System Calibration.} To guarantee high fidelity, the simulation engine is empirically anchored against discrete operating points and macroscopic scale-out trends measured from commercial hardware platforms.

\textbf{Final Output.} 
Upon completing the exploration budget, \prjname{} outputs the optimized, physically feasible XHS candidate architecture $\hat{h}$, alongside its customized workload mapping strategies and telemetry, 
tailored to the user's targeted set of applications.

\subsection{Decoupled Two-Level Optimization Strategy}
A na\"{i}ve approach to exploring the XHS design space would attempt to simultaneously optimize both the hardware architecture parameters and the software execution mapping. However, this joint search space is computationally intractable, consistent with the combinatorial growth observed in heterogeneous task scheduling~\cite{topcuoglu2002heft,arabnejad2014peft,parashar2019timeloop}.

We use a \textit{fiber-bundle-inspired view}\footnote{In mathematics, a fiber bundle~\cite{steenrod1999topology} can be viewed as a collection of spaces attached to the points of another space: every point in the base space carries its own fiber, and the collection of all such fibers forms a structured whole.} to organize this dependency. The base space represents the discrete hardware design space $H$. For every specific hardware configuration $h \in H$, the associated mapping space $S_h$ represents the possible software mappings, including tensor partitioning, scheduling, and routing choices. Because $S_h$ fundamentally depends on the resources and topology exposed by $h$, \prjname{} evaluates the candidates using a nested objective:
\begin{equation}
    \widetilde{\mathcal{O}}(h)
    =
    \max_{s \in S_h} \mathcal{O}(h, s),
    \qquad
    \hat{h}
    \in
    \operatorname*{arg\,max}_{h \in H}
    \widetilde{\mathcal{O}}(h).
\end{equation}
This mathematical expression serves as a search organization framework rather than a claim that the implementation exhaustively solves the full XHS design space; in practice, \prjname{} relies on heuristic search for the inner loop and reinforcement learning for the outer loop.

Guided by this organization, \prjname{} avoids intractable monolithic search by employing a \textbf{Decoupled Two-Level Exploration} strategy:
\begin{itemize}[leftmargin=*]
    \item \textbf{Inner Loop (Software Mapping Tuning):} Given a fixed hardware architecture $h$ proposed by the outer loop, the \textit{Mapper} heuristically searches the mapping space $S_h$ to select a high-quality execution mapping $s$. It adjusts tensor tiling, concurrency scheduling, and network routing to maximize performance on that specific topology.
    \item \textbf{Outer Loop (Hardware Architecture Evolution):}  Receiving the evaluated mapping performance via the \textit{Simulator}, the \textit{Optimizer} proposes updated XHS architectures $h \in H$. It iteratively adjusts the heterogeneous node ratio, memory pooling capacity, and multi-plane interconnect topology through local graph edits, strictly sampling only those candidates that pass the physical constraint verifier.
\end{itemize}

\subsection{Hardware-Validated Calibration Path}
Event-driven simulation must be anchored to measured system behavior before it can be trusted for cross-layer XHS-scale  projection. Prior simulation frameworks similarly rely on validated models to keep large-system studies meaningful~\cite{casanova2014simgrid,karandikar2018firesim,rashidi2020astrasim}. To prevent the framework from generating physically detached rankings, \prjname{} introduces a hardware-validated calibration path that serves two critical roles:

\textbf{1) Discrete-Point Anchoring.} \prjname{} first calibrates the simulator at a set of discrete operating points on heterogeneous testbeds. By matching observable quantities (e.g., protocol serialization overhead, effective link bandwidth, and tail latency~\cite{liu2025cxl,dean2013tail}) to actual hardware, this step grounds the local event-delay parameters to physical realities rather than nominal specifications.

\textbf{2) Scale-Out Trend Guardrail.} Second, \prjname{} verifies the simulator's behavior against real-system scale-out 
trends. By ensuring the simulated scaling curves align with physical measurements regarding congestion onset and queueing amplification, we establish a high-fidelity guardrail. This combined 
basis ensures that the simulator reliably evaluates large-scale XHS  candidate topologies that cannot be built directly during the exploration phase. Detailed calibration methodologies and results are provided in Section 6 and 7.

%% file: sec/04_modeling.tex
\section{Hardware Modeling}
\label{sec:hardware_modeling}

For systematic exploration, \prjname{} abstracts the vast XHS design space into a formal graph-based representation. Detailed catalogs, templates, and formalisms are in Appendix~\ref{app:appendix_modeling}.

\subsection{Multi-Layered Graph Representation}
\prjname{} treats each candidate hardware architecture $h$ as a stack of typed physical graphs across $4$ hierarchical layers:
\begin{equation}
    h = (G_1, G_2, G_3, G_4), \qquad G_k = (V_k, E_k, \lambda_k, \rho_k).
    \label{eq:graph_stack}
\end{equation}
Here, $G_k$ is the graph at physical layer $L_k$ ($k \in \{1, 2, 3, 4\}$), where $V_k$ contains hardware entities, $E_k$ contains links, $\lambda_k$ assigns hardware types, and $\rho_k$ stores physical attributes. The layers are:
1) \textit{Package-level} ($L_1$): heterogeneous dies and interfaces.
2) \textit{Node-level} ($L_2$): multi-tier memory and board routing.
3) \textit{Rack-level} ($L_3$): memory pools and intra-rack switches.
4) \textit{Cluster-level} ($L_4$): scale-out networks.

Each layer's graph space is generated from a Cartesian product:
\begin{equation}
    \mathcal{D}_{L_k} = \mathcal{N}_{L_k} \times \mathcal{I}_{L_k} \times \mathcal{T}_{L_k} \times \Theta_{L_k},
    \label{eq:design_space}
\end{equation}
where $\mathcal{N}_{L_k}$ is the component inventory, $\mathcal{I}_{L_k}$ is the interconnect medium, $\mathcal{T}_{L_k}$ is the topology template, and $\Theta_{L_k}$ contains template parameters. 

To interface with the mapper and simulator, each vertex $v \in V_k$ and edge $e \in E_k$ exposes a metric vector:
\begin{equation}
\begin{aligned}
    \rho_k(x) = \langle
    &C_{\text{peak}}, B_{\text{max}}, L_{\text{base}}, O_{\text{proto}}, \\
    &P_{\text{idle}}, P_{\text{active}}, Cost_{\text{rcu}}, U_{\text{space}}
    \rangle.
\end{aligned}
    \label{eq:metric_vector}
\end{equation}
Here, $x$ denotes an entity or link. $C_{\text{peak}}$ is peak compute throughput, $B_{\text{max}}$ is memory/link bandwidth, $L_{\text{base}}$ is base latency, and $O_{\text{proto}}$ captures protocol overheads. $P_{\text{idle}}$ and $P_{\text{active}}$ record power, $Cost_{\text{rcu}}$ denotes relative cost, and $U_{\text{space}}$ records physical footprint. \prjname{} uses neutral values for inapplicable dimensions and relies on $\lambda_k(x)$ for interpretation. 

\subsection{Physical Constraint Filtering}
A syntactically generated graph from $\prod \mathcal{D}_{L_k}$ is often physically invalid. To avoid wasting mapping and simulation cycles, \prjname{} applies a \textit{Physical Constraint Verifier}.

First, it enforces structural composition rules. For $k > 1$, a node must be a native component or a valid subgraph from the preceding layer:
\begin{equation}
    V_k \subseteq \mathcal{C}_{L_k} \cup \{G_{k-1}^{(i)} \mid G_{k-1}^{(i)} \in H_{k-1,\text{valid}}\}.
    \label{eq:composition_rule}
\end{equation}
Edges must connect compatible ports, bounded by the selected topology template.
    
Second, it evaluates systemic predicates against physical constraints, restricting the space to feasible points $H_{\text{valid}}$: 
\begin{equation}
\begin{aligned}
    \mathcal{D}_{\text{all}}
    &= \prod_{k=1}^{4} \mathcal{D}_{L_k}, \\
    H_{\text{valid}}
    &= \left\{ h \in \mathcal{D}_{\text{all}} \mid
    \Phi_{\text{feasible}}(h) \right\}, \\
    \Phi_{\text{feasible}}(h)
    &= \Phi_{\text{budget}}(h)
    \wedge \Phi_{\text{power}}(h)
    \wedge \Phi_{\text{thermal}}(h) \\
    &\quad
    \wedge \Phi_{\text{radix}}(h)
    \wedge \Phi_{\text{space}}(h).
\end{aligned}
    \label{eq:constraint_predicates}
\end{equation}
$\Phi_{\text{feasible}}$ is true only when all checks pass. These predicates filter configurations violating budget ($\Phi_{\text{budget}}$), power/cooling envelopes ($\Phi_{\text{power}}, \Phi_{\text{thermal}}$), switch port limits ($\Phi_{\text{radix}}$), and volumetric/cabling constraints ($\Phi_{\text{space}}$). Only valid candidates $h \in H_{\text{valid}}$ proceed to the optimization engine.

%% file: sec/05_cod.tex
\section{Decoupled Two-Level Optimization Engine}
\label{sec:optimization_engine}

The core of \prjname{} is a dual-loop optimization engine that addresses the  topology-mapping deadlock identified in 
Section~\ref{sec:background_motivation}. For any feasible hardware point $h$, the engine first searches its unique mapping space $S_h$ via an inner-loop \textit{Mapper} and evaluates the resulting trace via a \textit{Simulator}. The outer-loop \textit{Optimizer} then utilizes the evaluated performance to propose evolved hardware candidates.

\begin{figure}[!t]
    \centering
    \includegraphics[width=0.8\linewidth]{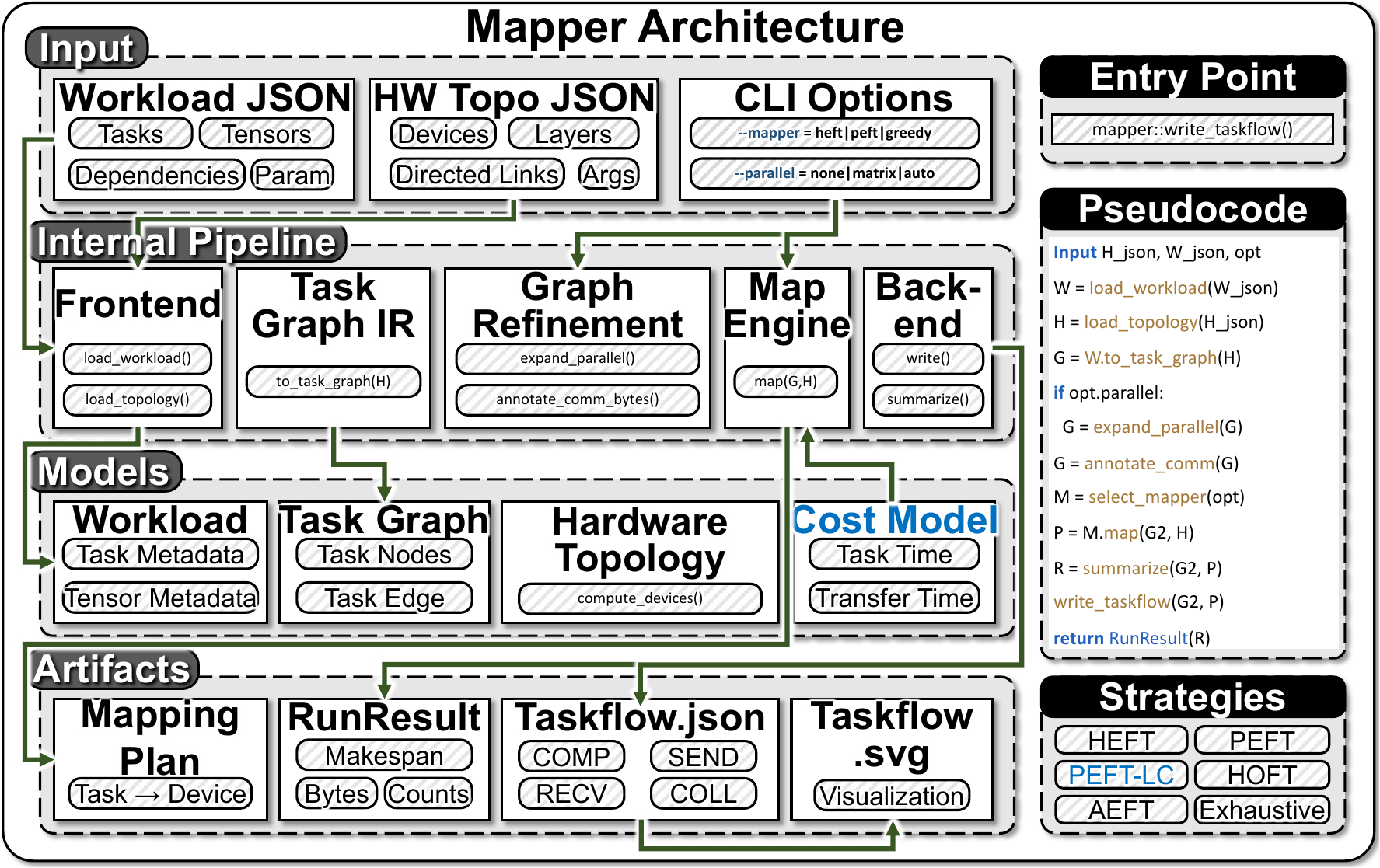}
    \caption{Overall architecture of the  workload mapper.
        }
    \label{fig:mapper_arch}
\end{figure}

\subsection{Graph-Based Workload Mapper}

The Mapper translates a hardware-independent workload DAG into a hardware-aware event trace for an XHS candidate. It annotates the logical DAG with communication volumes and heuristically selects valid placements and schedules bounded by the topology's resources.

Figure~\ref{fig:mapper_arch} summarizes the mapper architecture. It takes workload, hardware topology, and mapping options as inputs. The frontend parses these to construct a task-graph IR. The graph-refinement stage optionally expands parallel tasks and annotates edges with communication sizes. The map engine uses a topology-aware cost model to estimate task execution and transfer times. Based on these estimates, it produces a mapping plan using algorithms like HEFT~\cite{topcuoglu2002heft}, PEFT~\cite{arabnejad2014peft}, HOFT~\cite{10.1007/978-3-030-50371-0_1}, AEFT~\cite{WANG2025107576}, Greedy, or exhaustive search. We also introduce an optimized PEFT algorithm to handle heterogeneous costs efficiently. Finally, the backend writes the mapped event trace.

The mapper avoids exhaustive search by using list-scheduling heuristics. For large graphs, it optimizes PEFT scheduling overhead via cached cost estimates and locality-pruned device selection, whose effectiveness is evaluated in the experimental section.

\subsection{Hardware-Calibrated Heterogeneous-System Simulator}
\begin{figure}[!t]
    \centering
    \includegraphics[width=0.8\linewidth]{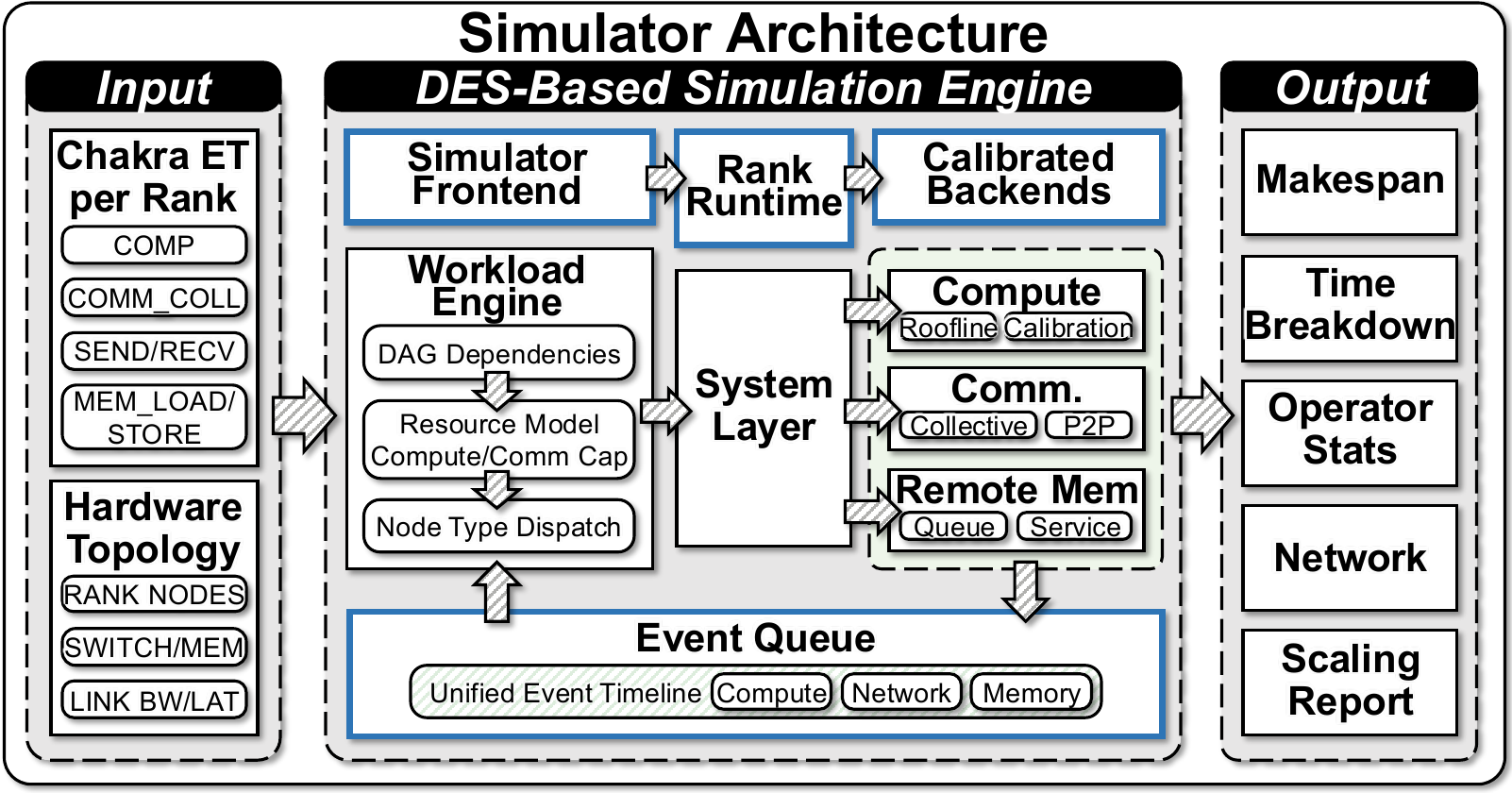}
    \caption{Overall architecture of the real-hardware-calibrated heterogeneous-system simulator.}
        \label{fig:simulator_arch}
\end{figure}

The hardware-aware trace is evaluated by a calibrated event-driven simulator. As shown in Figure~\ref{fig:simulator_arch}, it advances compute, point-to-point, collective, and remote-memory events on a unified timeline. This preserves causal dependencies and exposes resource contention across compute devices, network fabrics, and memory providers.

\textbf{Real-Hardware Calibration.} To ensure physically meaningful rankings, the simulator avoids relying solely on nominal specifications. Compute and communication timings are anchored by replay measurements from commercial systems and applied as calibrated performance models. Compute events combine measured operator behavior with roofline-style bounds~\cite{williams2009roofline}, retaining analytical coverage for unseen candidates while remaining grounded in physical reality.

\textbf{Explicit Heterogeneous Topology and Congestion Awareness.} Unlike template-based simulators, \prjname{} models candidates as explicit directed hardware graphs containing heterogeneous compute, switch, and memory nodes connected by asymmetric links. Communication events route over this topology, consuming shared resources. Consequently, serialization, queueing, and hotspot contention naturally emerge, allowing rack- and cluster-level congestion to directly impact the makespan.

\textbf{Remote-Memory Modeling.} \prjname{} treats remote-memory access as a first-class event rather than a fixed compute penalty. Remote accesses traverse the network, queue at the memory provider, and consume provider bandwidth. This isolates network latency from memory contention, yielding actionable telemetry for the outer optimizer. Ultimately, the simulator returns runtime, utilization, and bottleneck signals to the outer-loop \textit{Optimizer}, which ranks the candidates. After the exploration budget is exhausted, \prjname{} reports the final desired architecture $\hat{h}$ with its mapping and telemetry.

\subsection{Hardware Architecture Optimizer}
\label{sec:optimizer}

\begin{figure}[!t]
    \centering
    \includegraphics[width=0.8\linewidth]{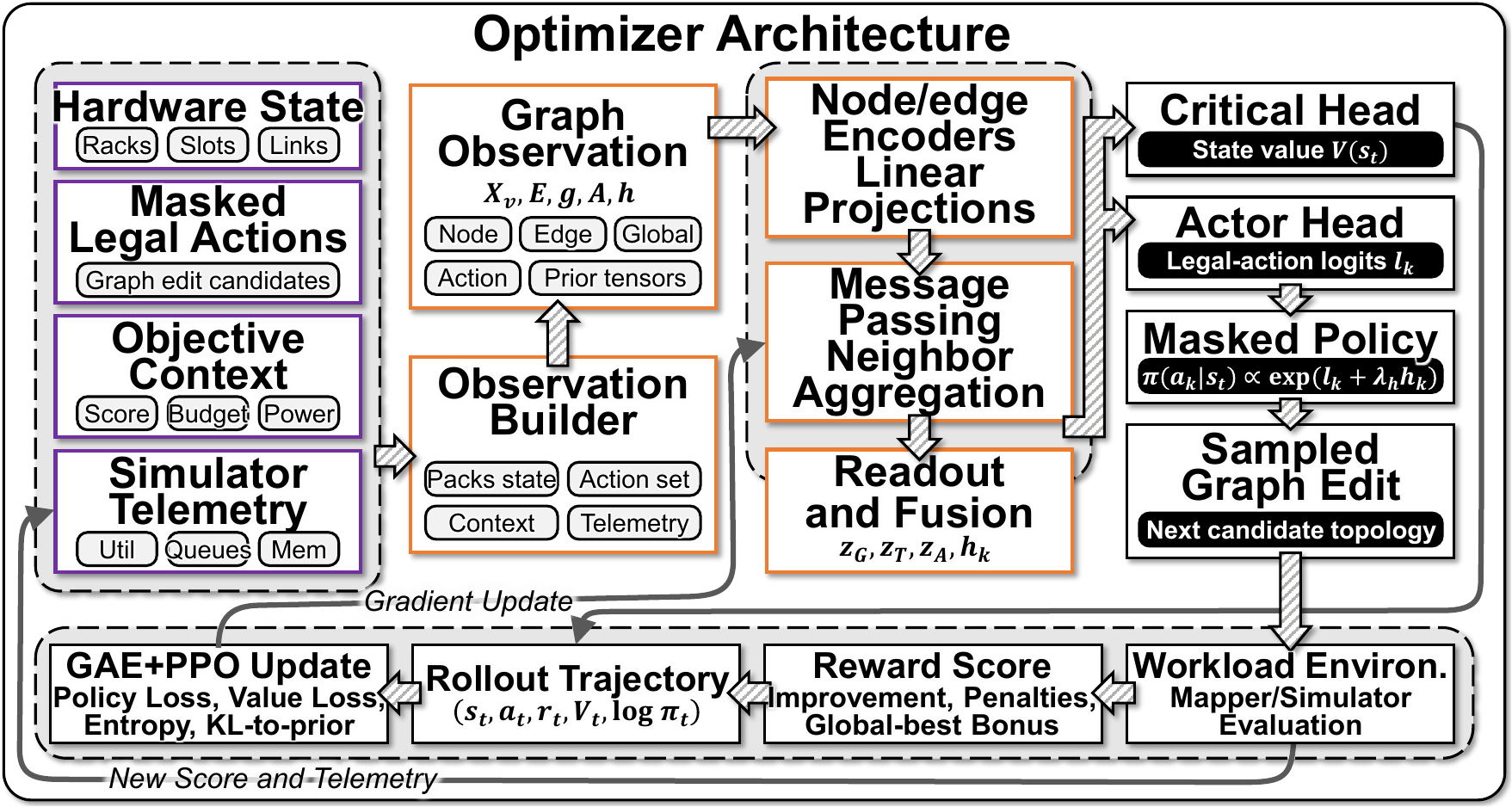}
    \caption{Architecture of the hardware optimizer. The GNN policy encodes the current hardware graph, combines learned action logits with simulator-derived bottleneck telemetry, applies legality masks to prune physically invalid graph edits, and sends valid successor architectures back to the mapper--simulator loop for reward evaluation.
        }
    \label{fig:optimizer_arch}
\end{figure}

To navigate the massive outer-loop hardware space, \prjname{} employs a Graph Neural Network (GNN)-based Reinforcement Learning (RL) agent. Instead of treating XHS design as a black-box environment, \prjname{} introduces three system-specific improvements to algorithms like TG-RL~\cite{zhou2020gnnreview, schulman2017ppo} to ensure convergence and feasibility. 
Figure~\ref{fig:optimizer_arch} summarizes the optimizer. At each step, it observes the hardware graph and simulator feedback, masks constraint-violating actions, and scores remaining actions.
The selected edit produces a candidate architecture, evaluated by the mapper and simulator, providing reward and telemetry for the next step.

\textbf{Legality-Constrained Action Masking.} To guarantee feasibility and maximize sample efficiency, \prjname{} projects physical constraint predicates (Eq.~\ref{eq:constraint_predicates}) into the action space. Before each decision, a legality mask prunes topological edits (e.g., node insertion, rewiring) violating power, port, or budget limits. This prevents wasting simulation budget on impossible architectures.

\textbf{Bottleneck-Guided Telemetry Prior.} To accelerate convergence, \prjname{} injects system intelligence into the actor network. We use simulator telemetry (Section 5.2) to construct a heuristic prior $h(o_t, a)$. Rather than relying solely on the GNN's logits $g_\theta(o_t, a)$, the action distribution combines the policy with this prior:
\begin{equation}
    \pi_\theta(a \mid o_t) = \text{softmax}_{a \in \mathcal{A}_t^{\text{valid}}} \left( g_\theta(o_t, a) + \lambda h(o_t, a) \right).
    \label{eq:telemetry_prior}
\end{equation}
If telemetry flags remote-memory contention, the prior biases actions toward adding memory capacity or link bandwidth. The coefficient $\lambda$ decays to balance initial heuristics with late-stage RL exploitation.

\textbf{Workload-Normalized Reward Stabilization.} Optimizing for a workload suite by aggregating raw runtimes destabilizes the RL gradient, as execution times span orders of magnitude (hours for LLMs vs. milliseconds for HPC traces). \prjname{} addresses this by transforming the runtime cost into a \textit{workload-normalized log-improvement} reward. This balances performance scaling (via geometric mean speedup) and penalizes over-budget trajectories. 

GNN state embeddings, reward functions, and PPO hyperparameters are detailed in Appendix~\ref{app:optimization_details}.

%% file: sec/06_calib.tex
\section{Real-System Calibration \& Scale Projection}
\label{sec:calibration}

\begin{table}[t]
\centering
\footnotesize
\setlength{\tabcolsep}{3pt}
\caption{Calibration and projection coverage. The first five rows are the architecture systems used for calibration across three anonymized commercial platforms; the final row is held out for projection validation. Accuracy distributions are reported separately in Figure~\ref{fig:calibration_error}. 
 }
\label{tab:calibration_configurations}
\begin{tabular}{@{}lllll@{}}
\toprule
\textbf{Platform} & \textbf{System} & \textbf{Scale} & \textbf{Fabric} & \textbf{Use} \\
\midrule
Platform A & A800-class & 2--16 GPU & NVLink \& IB & Calibration \\
Platform A & H800-class & 2--8 GPU  & NVLink & Calibration \\
Platform A & H20-class  & 2--8 GPU  & NVLink & Calibration \\
Platform B & RTX4090-class & 2 GPU & PCIe & Calibration \\
Platform C & C500-class & 1 GPU & PCIe & Calibration \\
\midrule
Held-out & L20-class & 8 GPU & NVLink & Projection \\
\bottomrule
\end{tabular}
\end{table}

Reliable architectural comparison requires that the simulator correctly orders candidate designs rather than reproducing cycle-exact behavior. \prjname{} treats calibration as a guardrail: the uncalibrated simulator (Section~\ref{sec:optimization_engine}) provides roofline- and serialization-based upper bounds on event service times, and calibration fits efficiency corrections to match measured hardware. Corrections are fit on five architecture systems from three commercial platforms and projected onto larger unbuilt XHS candidates.

\subsection{Calibration Methodology}

For each compute operator $o$, the simulator derives a roofline upper bound~\cite{williams2009roofline} $\max(u^{\mathrm{flop}}_o, u^{\mathrm{mem}}_o)$ from peak throughput and memory bandwidth. Calibration introduces three knobs per operator---FLOP efficiency $\phi_o$, bandwidth efficiency $\beta_o$, and launch overhead $\ell_o$---yielding $\hat{t}_o = \ell_o + \max(u^{\mathrm{flop}}_o/\phi_o,\; u^{\mathrm{mem}}_o/\beta_o)$. Communication events are handled analogously: uncalibrated link-serialization time is corrected by protocol-startup latency and bandwidth-derating factors, with separate parameters for point-to-point and collective transfers.

Ground-truth timings come from calibration workloads processed by the Mapper into hardware-aware event traces. The trace specifies kernel launches, tensor movements, and collective events to replay on the physical platform to measure event durations with warmup~\cite{sridharan2026mlcommonschakra}. Since roofline ceilings are independent of calibration knobs, a single simulator pass produces all reference values, and each knob is recovered via weighted least-squares regression. Residuals are weighted by $1/t^2$ to minimize relative error, preventing large kernels from dominating. Efficiencies are clamped to $(0,1]$. When a single parameter set cannot cover all operator sizes, the calibrator splits the size axis into contiguous segments if it yields a statistically significant improvement.

We separate hardware coverage from calibration accuracy: Table~\ref{tab:calibration_configurations} reports the platforms used to fit and validate profiles, while Figure~\ref{fig:calibration_error} details the calibration-error distribution.

\subsection{Hardware Coverage}
As shown in Table~\ref{tab:calibration_configurations}, our dataset covers five architecture systems from three commercial platforms. These include NVLink-based scale-up hosts, PCIe-based hosts, and a multi-node configuration connected via 4$\times$200\,Gbps InfiniBand, spanning intra-machine NVLink/PCIe and inter-machine RDMA paths. The held-out L20-class platform evaluates whether fitted profiles can project to unseen systems.

For each platform, we run workloads stressing compute and communication---sparse iterative solvers for irregular memory access and point-to-point transfers, and dense LLM forward passes for collective bandwidth. The replay traces and simulator-predicted upper bounds provide the observations to fit calibration knobs per machine.

\subsection{Scale-Out Projection}

Per-machine profiles alone cannot cover the full XHS design space, as simulators may mispredict large-scale congestion or synchronization overheads~\cite{casanova2014simgrid,karandikar2018firesim}. \prjname{} therefore pools the fitted profiles across platforms and treats each knob as a regression target over hardware attributes: peak throughput, memory/link bandwidth, link latency, device count, and hop-distance. Coefficients are fit via ridge regression, weighting each sample by its anchoring quality.

Given any candidate $h \in H_{\text{valid}}$, the regression model maps its hardware attributes to a complete set of calibration knobs, projecting profiles for unbuilt configurations. We reserve the L20-class 8-GPU NVLink platform as a held-out system to validate this projection model. Detailed L20 projection errors are reported in Section~\ref{sec:evaluation}.

%% file: sec/07_eval.tex
\section{Evaluation}
\label{sec:evaluation}

\providecommand{\caseinsight}[1]{\begin{center}
{\setlength{\fboxsep}{5pt}\fbox{\begin{minipage}{0.95\linewidth}
\small \textbf{Take Away.} #1
\end{minipage}}}
\end{center}}

\begin{table}[t]
  \centering
  \scriptsize
  \setlength{\tabcolsep}{2.0pt}
  \renewcommand{\arraystretch}{1.05}
  \caption{HPCG workload configurations. NNZ denotes the number of nonzeros, and Iter. denotes iterations.}
  \label{tab:hpcg-workloads}
  \begin{tabular*}{0.9\columnwidth}{@{\extracolsep{\fill}}cccccc@{}}
    \toprule
    Workload & Grid & Rows & NNZ & Iter. & Tasks \\
    \midrule
    \texttt{n16\_i3}    & $16^3$  & 4,096     & 97,336     & 3   & 131 \\
    \texttt{n32\_i5}    & $32^3$  & 32,768    & 830,584    & 5   & 217 \\
    \texttt{n64\_i50}   & $64^3$  & 262,144   & 6,859,000  & 50  & 2,152 \\
    \texttt{n104\_i100} & $104^3$ & 1,124,864 & 29,791,000 & 100 & 4,302 \\
    \texttt{n104\_i500} & $104^3$ & 1,124,864 & 29,791,000 & 500 & 21,502 \\
    \bottomrule
  \end{tabular*}
\end{table}

\begin{table}[t]
  \centering
  \scriptsize
  \setlength{\tabcolsep}{2.0pt}
  \renewcommand{\arraystretch}{1.05}
  \caption{Sparse direct solver workload configurations. NNZ denotes the number of nonzeros.}
  \label{tab:direct-solve-workloads}
  \begin{tabular*}{0.9\columnwidth}{@{\extracolsep{\fill}}cccccc@{}}
    \toprule
    Workload & Rows & NNZ &Tasks \\
    \midrule
    \texttt{apache2}    & 715,176     & 4,817,870     & 32,411 \\
    \texttt{CoupCons3D} & 416,800    & 17,277,420    & 32,206 \\
    \texttt{ecology1}   & 1,000,000   & 4,996,000  & 29,426 \\
    \bottomrule
  \end{tabular*}
\end{table}

\begin{table}[t]
  \centering
  \scriptsize
  \setlength{\tabcolsep}{2.0pt}
  \renewcommand{\arraystretch}{1.05}
  \caption{Model configurations for LLM workload. Req. Len. denotes the maximum request length.}
  \label{tab:llm-model-config}
  \begin{tabular*}{\columnwidth}{@{\extracolsep{\fill}}lcccccccc@{}}
    \toprule
    Model & Param. & Attn. & Type & $L$ & $H$ & $d$ & $d_{\mathrm{FFN/E}}$ & Req. Len. \\
    \midrule
    Qwen3.5  & 32B      & GQA & Dense & 64 & 24 & 5,120 & 17,408 & 262K \\
    Llama3.1 & 70B      & GQA & Dense & 80 & 64 & 8,192 & 28,672 & 128K \\
    GPT-3    & 13B      & MHA & Dense & 40 & 40 & 5,140 & 20,560 & 2K \\
    Gemma4   & 27B      & GQA & Dense & 60 & 32 & 5,376 & 21,504 & 262K \\
    Mixtral  & $8{\times}$7B & GQA & MoE & 32 & 32 & 4,096 & 14,336 & 32K \\
    \bottomrule
  \end{tabular*}
\end{table}

This section evaluates \prjname{} systematically across its core components and its end-to-end capability. 
The evaluation answers for four key questions: 
\textbf{Q1 (Mapper):} Does the hardware-aware mapper produce near-optimal schedules in tractable spaces, and is it competitive with other strong algorithms in large-scale scenarios?
\textbf{Q2 (Simulator):} How accurately does the calibrated event-driven simulator match real-system measurements, and how does its simulation time compare with ASTRA-SIM~\cite{rashidi2020astrasim}?
\textbf{Q3 (Optimizer):} Can the outer-loop search engine find near-optimal hardware architectures in small spaces, and does it outperform baseline strategies (e.g., Random, Greedy) in massive design spaces?
\textbf{Q4 (Case Studies):} Under realistic physical constraints, what workload-specific XHS architectures does the complete framework select for distinct application paradigms?

\subsection{Experimental Workload}

We evaluate two workload families stressing different architectural dimensions. Sparse workloads represent memory-sensitive HPC/graph computations with irregular access. LLM inference captures dense tensor computation with regular pipeline and tensor-parallel communication~\cite{narayanan2021megatron,zheng2022alpa}.

\textbf{Sparse computation suite.}
This suite contains two trace families: HPCG (Table~\ref{tab:hpcg-workloads})~\cite{doi:10.1177/1094342015593158} and sparse direct solvers (Table~\ref{tab:direct-solve-workloads})~\cite{10.1145/3774934.3786442}. HPCG represents iterative sparse linear solvers featuring irregular memory access, multigrid structures, tri- angular-solve dependencies, SpMV, and reductions. We generate these by converting local HPCG kernel traces into our workload representation.

To complement HPCG, we include GPU sparse direct solver traces following the Trojan Horse~\cite{10.1145/3774934.3786442} aggregate-and-batch model. These capture state-of-the-art scheduling in sparse factorization, where fine-grained tasks are prioritized, aggregated, and batch-dispatched via DAG dependencies.

\textbf{LLM inference suite.}
This suite (Table~\ref{tab:llm-model-config}) covers representative decoder-only and MoE models: Qwen3.5~\cite{qwen3.5}, Llama3.1~\cite{grattafiori2024llama3herdmodels}, GPT-3~\cite{NEURIPS2020_1457c0d6}, Gemma~\cite{gemmateam2024gemmaopenmodelsbased}, and Mixtral~\cite{jiang2024mixtralexperts}. They provide dense tensor computation with predictable shapes. Unlike irregular sparse DAGs, LLM inference exposes parallelism through batched matrix multiplications, attention, and structured pipeline/tensor-parallel communication.

Instantiated with a 2048 request length, model executions are converted into our common workload representation. The resulting traces stress accelerator throughput, memory bandwidth, and collective communication.

\begin{figure}[!t]
    \centering
    \includegraphics[width=0.8\linewidth]{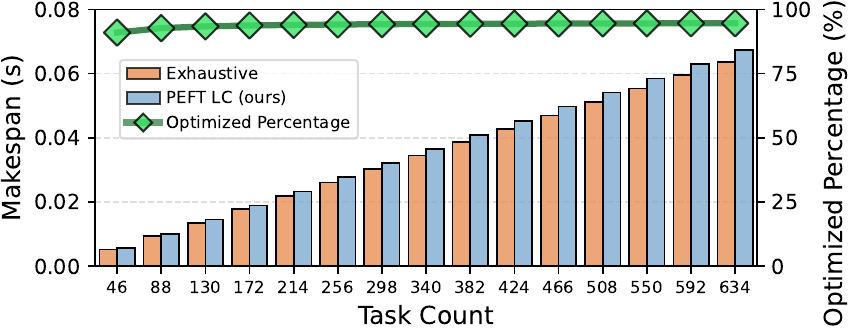}
    \caption{Mapper Near-Optimality:  Bar chart showing the relative performance gap between \prjname{}'s mapper and exhaustive optimal search on tractable hardware topologies.}
        \label{fig:mapper_near_opt}
\end{figure}

\begin{figure}[!t]
    \centering
    \includegraphics[width=0.89\linewidth]{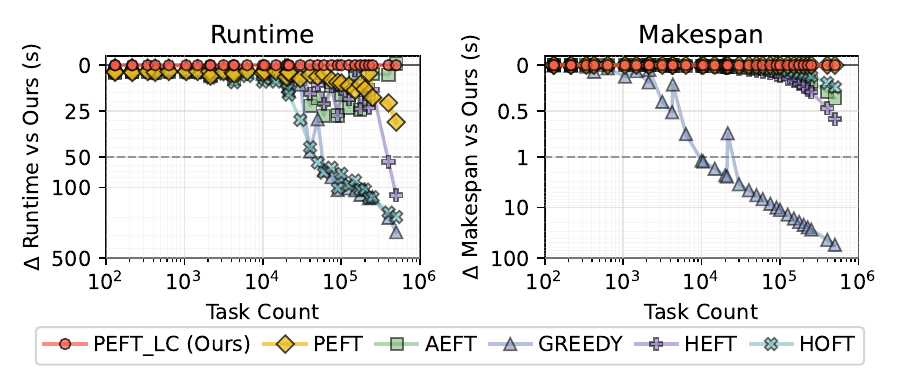}
    \caption{Mapper algorithm against other strong baseline algorithms: Runtime/Makespan comparison between CHASE Mapper and baselines (e.g., PEFT) on large-scale workloads.}
    \label{fig:mapper_sota}
\end{figure}

\subsection{Mapper Evaluation}
\label{sec:mapper-eval}

The mapper bridges workload DAGs and hardware topologies. We evaluate its quality through near-optimality on tractable spaces and mapping overhead on large workloads.

\textbf{Near-Optimality Validation on Tractable Spaces.}
We study small-scale workloads where the mapping space $\mathcal{S}_{h}$ under a fixed topology $h$ can be enumerated. This provides the exact optimal mapping via exhaustive search as a baseline.

On a 4-GPU ring topology, we evaluate 15 tractable HPCG workloads (46--634 tasks) by varying CG iterations. As Figure~\ref{fig:mapper_near_opt} shows, our PEFT-LC algorithm approaches the global optimum, averaging a 6.06\% gap (5.68\% weighted average) from exhaustive search. Although exhaustive enumeration is infeasible for larger workloads, this exact validation confirms the mapper’s topology-aware decisions remain near-optimal.

\textbf{Overhead Optimization at Scale.}
We evaluate mapping overhead on large DAGs where exhaustive search is intractable, comparing \prjname{} against HEFT~\cite{topcuoglu2002heft}, PEFT~\cite{arabnejad2014peft}, AEFT~\cite{WANG2025107576}, and HOFT~\cite{10.1007/978-3-030-50371-0_1}. As shown in Figure~\ref{fig:mapper_sota}, across 38 HPCG DAGs (130--500,014 tasks), \prjname{} reduces mapper runtime by 60.5\% on average versus PEFT while matching its estimated makespan. Compared to HEFT, AEFT, and HOFT, \prjname{} reduces mapper runtime by 63.6\%, 64.4\%, and 82.7\%, respectively, while improving median makespan by 0.096\%, 0.059\%, and 0.039\%. Thus, \prjname{} preserves PEFT's schedule quality while significantly reducing mapper overhead.

\begin{figure}[!t]
    \centering
        \includegraphics[width=0.8\linewidth]{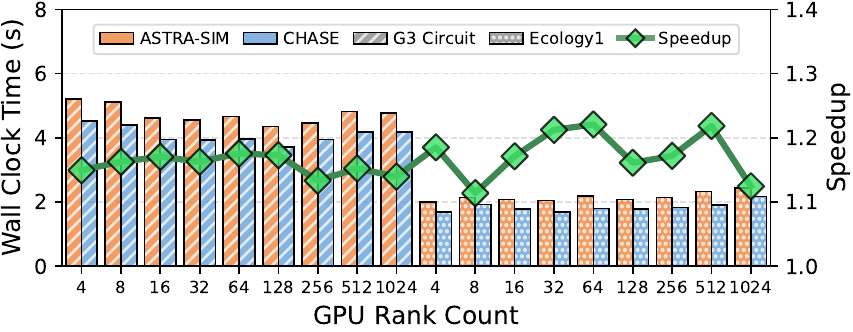}
    \caption{Simulator performance against ASTRA-SIM: wall-clock simulation time (seconds) under matched workload traces and hardware configurations.}
    \label{fig:sim_performance}
\end{figure}

\begin{figure}[!t]
    \centering
    \includegraphics[width=0.8\linewidth]{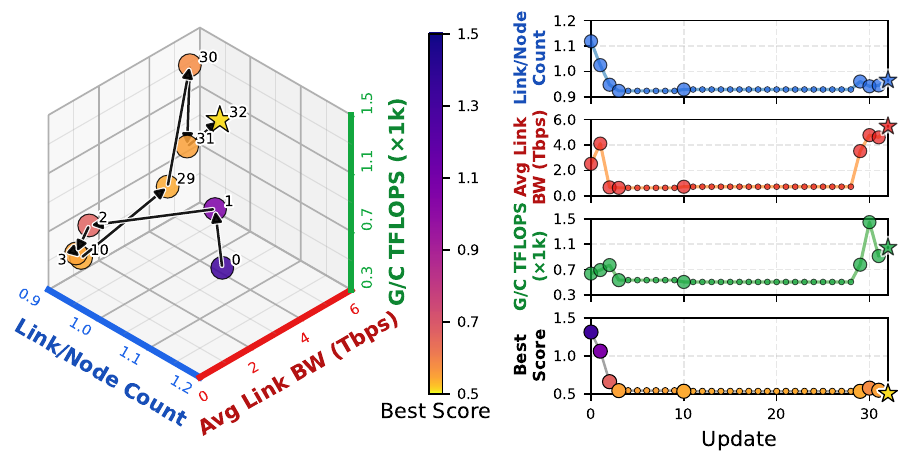}
    \caption{Optimization process. The outer-loop optimizer proposes physically valid hardware candidates, the mapper derives a topology-specific schedule for each candidate, the simulator returns performance and bottleneck telemetry, and the optimizer accordingly update the next candidate.         }
    \label{fig:optimize-process}
\end{figure}

\begin{figure*}[t]
    \centering
    \includegraphics[width=0.9\linewidth]{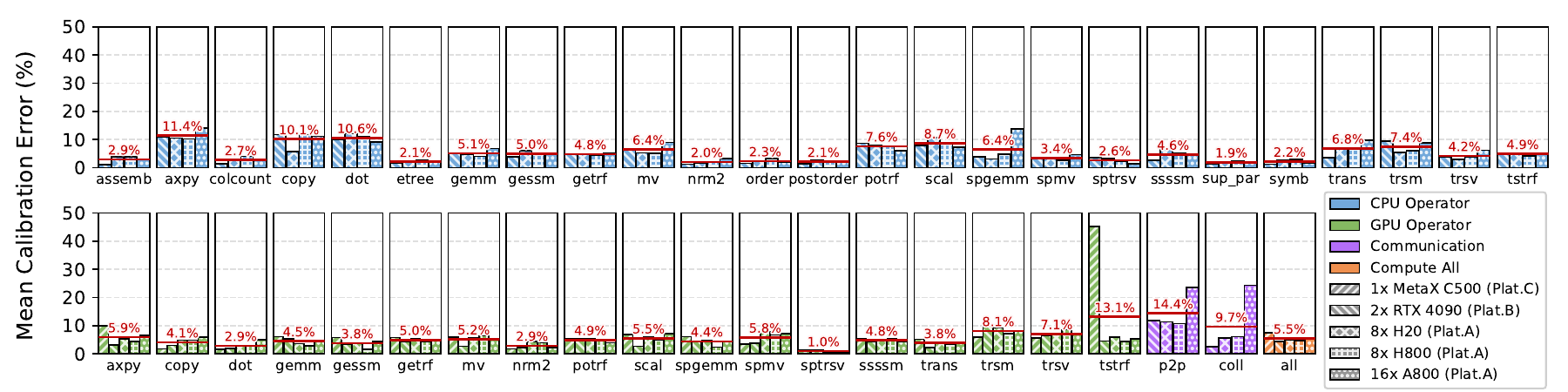}
    \caption{Mean calibration error of the \prjname{} simulator across different operators and hardware platforms compared to real-system ground truth. 
        C500 does not optimize \texttt{tstrf} kernel, leading to a outlier point.
    }
    \label{fig:calibration_error}
\end{figure*}

\subsection{Simulator Evaluation}

To ensure credible rankings, we evaluate the simulator's fidelity against real hardware and its efficiency for inner-loop DSE. We use ASTRA-SIM~\cite{rashidi2020astrasim} as the baseline, comparing wall-clock time under matched traces and hardware.

\textbf{Accuracy against Real-System Measurements.}
Following Section~\ref{sec:calibration}, we report the mean absolute relative error on each platform for compute and communication. Figure~\ref{fig:calibration_error} shows the per-operator error distribution across GPU types and interconnects. Table~\ref{tab:calibration_configurations} reports the platform coverage. 

Compute errors average 4.4--5.0\% across NVIDIA GPUs and 7.5\% on Metax C500. Intra-machine communication errors average below 10\% on NVLink and PCIe fabrics. To test generalization, we apply the trained regression model to a held-out NVIDIA L20 cluster. The 5.8\% mean error on L20 confirms accurate projection on unseen cases.

\textbf{Simulation-Time Comparison against ASTRA-SIM.}
We compare \prjname{}'s host-side wall-clock time against ASTRA-SIM~\cite{rashidi2020astrasim}. \prjname{} reduces simulation time by 15.5\% over ASTRA-SIM on large LLM traces and sustains $2.10 \times 10^5$ events per second, enabling extensive candidate evaluation (Figure~\ref{fig:sim_performance}).

\subsection{Optimizer Evaluation}

\begin{figure}[!t]
    \centering
    \includegraphics[width=0.8\linewidth]{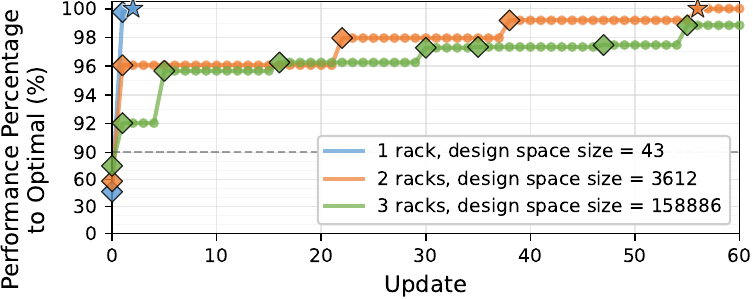}
        \caption{Optimizer Near-Optimality: Trajectory of the best-found architecture's performance gap to the true exhaustive optimum in a tractable finite hardware space.}
    \label{fig:optimizer_near_opt}
\end{figure}

The outer-loop TG-RL Optimizer is responsible for navigating the massive, physically constrained XHS design space. Figure \ref{fig:optimize-process} shows the Optimization process on a projected search space. We evaluate its search efficiency and optimality.

\textbf{Near-Optimality in Tractable Hardware Spaces.}
We restrict the hardware design space $\mathcal{H}$ to a small, finite set (e.g., 158,886 valid candidates) where exhaustive simulation is possible. Figure~\ref{fig:optimizer_near_opt} shows the gap between the optimizer-selected hardware $\hat{h}$ and the exhaustive best $h^{*}$ across search steps. On the tractable spaces, the TG-RL optimizer mitigates the risk of being trapped in local minima, consistently converging to near-global optima within 64 iterations.

\textbf{Search Efficiency in Massive Design Spaces.}
For practical XHS exploration, the space $\mathcal{H}$ is computationally intractable. We compare our TG-RL optimizer against standard search baselines: Random Search, Greedy Search, and Simulated Annealing. 
Under the same simulator evaluation budget, our TG-RL optimizer, guided by workload-normalized rewards and legality masking, reaches a stable high-quality architecture within 16 iterations. In contrast, the baseline methods remain below the TG-RL solution quality even after 64 iterations, demonstrating the higher sample efficiency of TG-RL in massive design-space exploration.

\subsection{Case Study 1: Sparse Computation}
\begin{figure}[!t]
    \centering
    \includegraphics[width=0.8\linewidth]{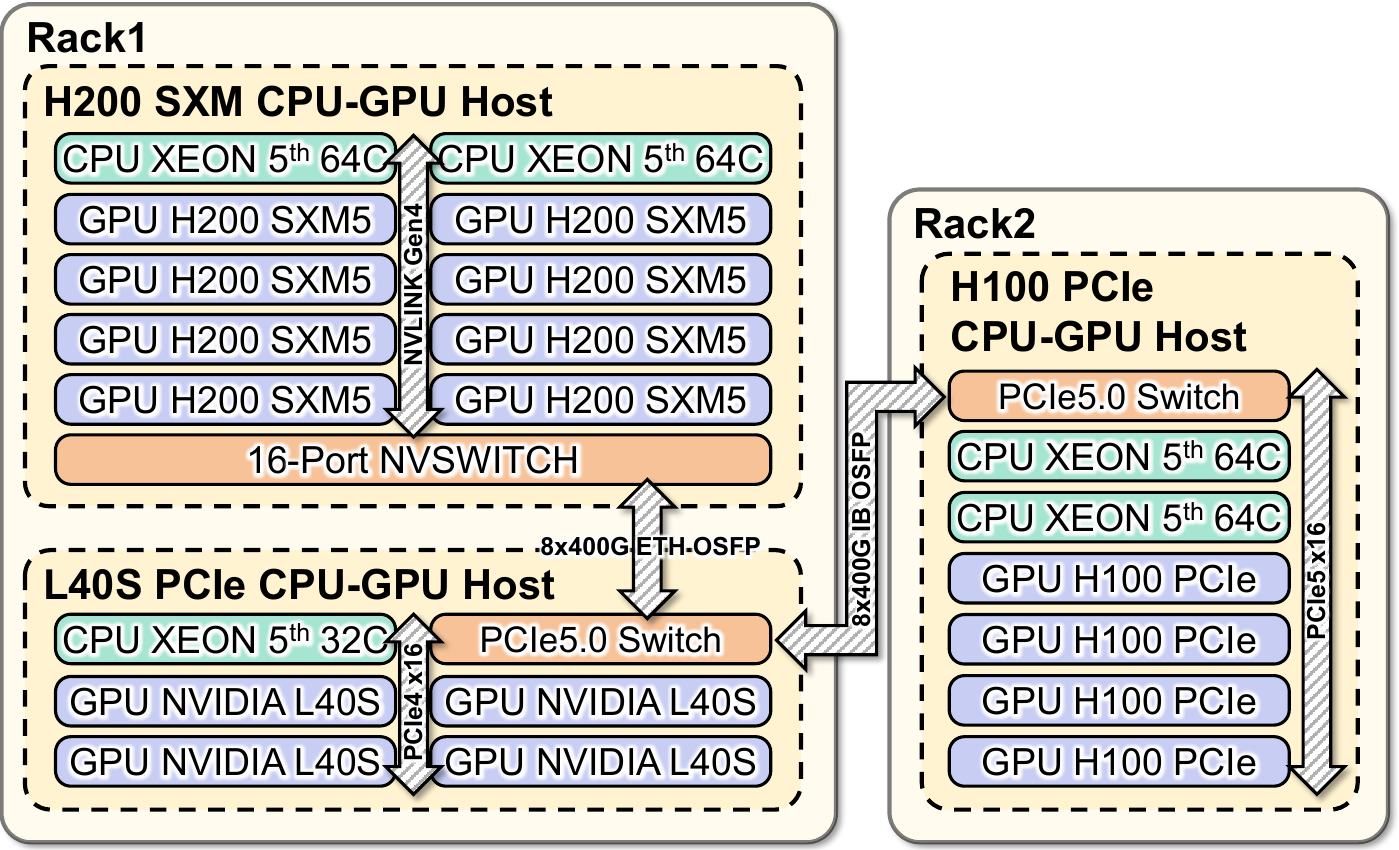}
    \caption{Sparse computation hardware architecture obtained through optimization.}
    \label{fig:sparse_architecture}
\end{figure}

We first examine the architecture selected for the sparse computation suite.
Sparse workloads expose irregular task DAGs with a mixture of dependency-critical factorization tasks and wider update phases, so the searched topology can reveal whether the optimizer simply prefers the fastest GPU host or differentiates resources by task role.
The selected architecture in Figure~\ref{fig:sparse_architecture} does not form a uniformly fastest-GPU pod.
It contains a high-bandwidth H200-based execution island together with H100-based CPU-GPU hosts.
This composition separates a high-performance tier from additional CPU-GPU hosts that provide parallel execution breadth under the same physical and cost constraints.

As shown in Table~\ref{tab:sparse_mapping_analysis}, we use three Trojan Horse~\cite{10.1145/3774934.3786442} factorization traces to interpret why this design pattern appears.\footnote{
These traces are used only for post-hoc mapping analysis in Table~\ref{tab:sparse_mapping_analysis}; we evaluate the complete workload suite during the sparse optimization run.}
The mapper places a large fraction of all analyzed tasks on the H200 host, and more importantly, it concentrates dependency-critical factorization operators on that host.
For example, the \texttt{getrf} operator is almost entirely mapped to the H200 host across all three traces, while \texttt{gessm} is also strongly concentrated there for \texttt{apache2} and \texttt{ecology1}.

\begin{table}[!t]
\centering
\scriptsize
\setlength{\tabcolsep}{2.0pt}
\renewcommand{\arraystretch}{1.05}
\caption{Analysis subset for the sparse case study. These Trojan-horse traces~\cite{10.1145/3774934.3786442} are selected from the full sparse optimization suite for post-hoc placement analysis; they are not the complete workload suite used by the optimizer.}
\label{tab:sparse_mapping_analysis}
\begin{tabular*}{0.8\columnwidth}{@{\extracolsep{\fill}}ccccc@{}}
\toprule
\textbf{Trace} & \textbf{Tasks} & \textbf{\texttt{getrf}} & \textbf{\texttt{gessm}} & \textbf{Spd.} \\
 & \textbf{on H200 tier} & \textbf{on H200 tier} & \textbf{on H200 tier} & \textbf{($\times$)} \\
 & \textbf{(\%)} & \textbf{(\%)} & \textbf{(\%)} & \\
\midrule
\texttt{CoupCons3D} & 60.3 & 100.0 & 77.0 & 3.93 \\
\texttt{apache2}    & 54.4 & 98.6  & 93.1 & 2.32 \\
\texttt{ecology1}   & 59.2 & 95.4  & 94.6 & 1.67 \\
\bottomrule
\end{tabular*}
\end{table}

This placement pattern matches the dependency structure of sparse factorization DAGs.
Panel and frontier operations, e.g., \texttt{getrf} and \texttt{gessm}, act as high-fanout producers for later update tasks, so accelerating them shortens the effective DAG frontier.
The H100-based hosts  absorb off-critical-path update work and provide additional CPU/GPU ranks without assigning every task to the most expensive host type.

\begin{table}[t]
\centering
\footnotesize
\setlength{\tabcolsep}{2.5pt}
\caption{Topology and performance summary for the sparse case study. Both systems are under the same cost and power constrains.}
\label{tab:sparse_case_study_summary}
\begin{tabularx}{\linewidth}{@{}lXX@{}}
\toprule
\textbf{Metric} & \textbf{El Capitan-like Baseline} & \textbf{\prjname{} Topology} \\
\midrule
GPU composition & 128 H100 SXM GPUs & 8 H200 + 4 H100 + 4 L40S\\
Host composition & 32 compute nodes & 1 HGX H200 + 1 HGX H100 + 1 L40S \\
CPU ranks & 64 & 5 \\
Geomean speedup & 1.00$\times$ & 6.20$\times$ \\
\bottomrule
\end{tabularx}
\end{table}

Table 8 shows the topology-level trade-off of the selected sparse-computation architecture against an El Capitan~\cite{osti_2584749}-like baseline, a representative state-of-the-art exascale system. Although the optimized topology uses fewer high-end GPU resources than the baseline, it achieves a higher geometric-mean speedup under the same cost and power constrains.
The resulting design suggests a simple allocation rule for sparse pods: use expensive high-bandwidth hosts where they shorten the DAG frontier, and use lower-cost heterogeneous hosts where additional execution breadth is sufficient.

\begin{tightbox}
{\textbf{Takeaway.} Sparse workloads favor criticality-aware heterogeneous pods rather than uniformly fastest-GPU pods: high-bandwidth resources should be concentrated on dependency-critical frontier operations, while cheaper CPU-GPU hosts provide parallel breadth for update work.}
\end{tightbox}

\subsection{Case Study 2: LLM Workloads}
The LLM case study exposes a different scale-up pattern. Dense transformer inference features regular GPU-resident computation and structured collectives. Thus, the main design question is how to align model parallelism with hardware scale-up boundaries, rather than placing irregular tasks.

\begin{figure}[!t]
    \centering
    \includegraphics[width=0.8\linewidth]{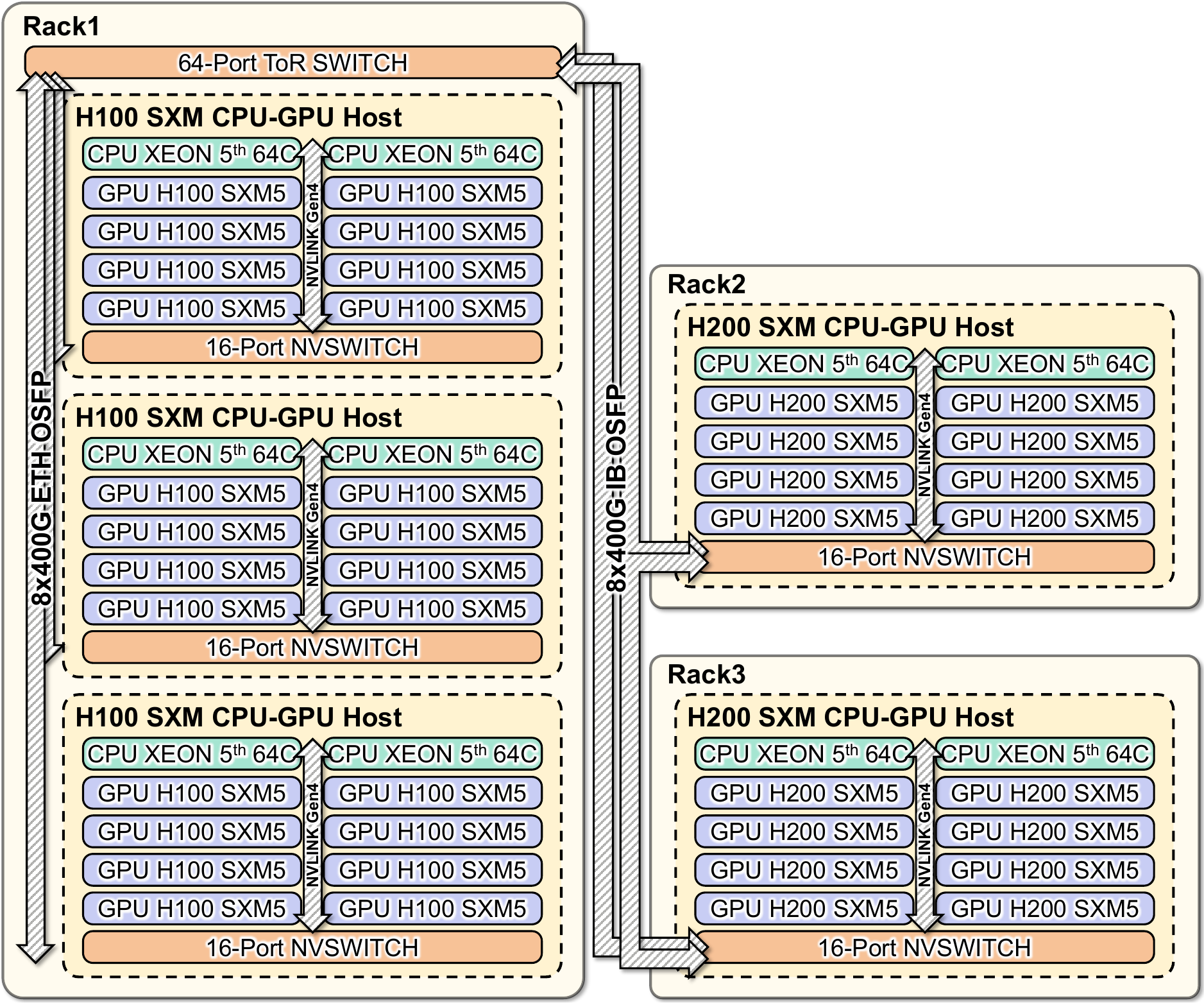}
    \caption{LLM-specific hardware architecture obtained through optimization.}
    \label{fig:llm_searched_architecture}
\end{figure}

\begin{table}[t]
\centering
\footnotesize
\setlength{\tabcolsep}{1.5pt}
\caption{Topology and performance summary for the LLM case study. Both systems are under the same constrains.}
\label{tab:llm_case_study_summary}
\begin{tabularx}{\linewidth}{@{}lll@{}}
\toprule
\textbf{Metric} & \textbf{NVL72-like Baseline} & \textbf{\prjname{} Topology} \\
\midrule
GPU composition & 72 H100 SXM GPUs & 16 H200 SXM + 24 H100 SXM \\
Host composition & 18 compute trays & 2 HGX H200 + 3 HGX H100 \\
CPU ranks & 36 & 10 \\
Geomean speedup & 1.00$\times$ & 2.12$\times$ \\
\bottomrule
\end{tabularx}
\end{table}

As Table~\ref{tab:llm_case_study_summary} summarizes, \prjname{} selects five HGX-style 8-GPU hosts (two H200, three H100), totaling 40 GPUs and 10 CPUs. This topology improves geomean performance by 2.12$\times$ and reduces GPU count, cost, and power compared to the 64-GPU H100 SuperPOD baseline (Table~\ref{tab:llm_case_study_summary}).

\textbf{Regular scale-up islands.}
The LLM topology comprises regular 8-GPU scale-up islands connected by rack-level fabrics. This matches dense transformer execution, where GPU throughput, HBM bandwidth, and low-latency GPU-GPU interconnects are critical. Unlike the sparse design, \prjname{} avoids CPU-only or PCIe-attached hosts; CPUs do not shorten the critical path, and PCIe links weaken tensor-parallel communication. H200/H100 heterogeneity appears at the host level: H200s serve bandwidth-sensitive stages, while H100s provide cost-effective dense compute.

\textbf{Collectives remain local to scale-up islands.}
The LLM traces generate significant collective traffic (e.g., 7.0\,GB for Llama, 4.4\,GB for Qwen/Gemma). The optimized topology changes where this communication occurs rather than removing it. For Qwen, Llama, Gemma, and Mixtral, the mapper selects tensor parallelism $\mathrm{TP}=8$ and pipeline parallelism $\mathrm{PP}=5$, fitting each TP group within one 8-GPU HGX host. Consequently, intra-host NVSwitch serves allreduce, allgather, and MoE collectives, restricting inter-host traffic to pipeline boundaries. The scale-out fabric acts as a pipeline network, not a primary collective substrate.

\textbf{More GPUs do not always improve LLM performance.}
A use-all-GPU policy on the 64-GPU baseline induces overly fine tensor parallelism. For Llama, the baseline chooses $\mathrm{TP}=64$ and $\mathrm{PP}=1$ (25,729 tasks). The optimized 40-GPU topology induces $\mathrm{TP}=8$ and $\mathrm{PP}=5$ (3,249 tasks), improving runtime by 1.20$\times$. Similarly, Gemma shifts from $\mathrm{TP}=32,\mathrm{PP}=2$ to $\mathrm{TP}=8,\mathrm{PP}=5$, improving runtime by 1.72$\times$, and Mixtral achieves a 1.13$\times$ speedup, where excessive tensor parallelism fragments operators and amplifies overheads. The superior design provides a GPU count and boundaries matching useful model-parallel granularity without the largest pool.

Table~\ref{tab:llm_case_study_summary} compares this optimized topology against a commercial SuperPOD baseline. Overall, tensor-parallel groups should fit inside high-bandwidth scale-up islands, pipeline parallelism should connect these islands, and GPU count must match useful parallelism granularity rather than be maximized blindly.

\begin{tightbox}
{\textbf{Takeaway}. Dense LLM workloads favor model-parallelism-aware scale-up islands: tensor-parallel groups should stay within high-bandwidth GPU domains, while pipeline parallelism connects those domains without forcing the system to use the largest possible GPU pool.}
    
\end{tightbox}

%% file: sec/08_con.tex
\section{Conclusion}

\prjname{} explores XHS as a coupled hardware-mapping problem, modeling architectures as hierarchical typed graphs, pruning via physical constraints, and evaluating through a mapper, simulator, and optimizer. Achieving near-exhaustive mapping optima and near-global optima across 158,886 candidates, it matches production compute behavior and reveals that sparse workloads favor criticality-aware pods while dense LLMs favor model-parallelism islands. XHS architectures should therefore be selected from workload dependency and physical constraints rather than homogeneous SuperPOD templates or isolated specs.

%% file: sec/10_app.tex
\section{Detailed Hardware Description Space Modeling}
\label{app:appendix_modeling}

This appendix specifies the hardware description space $\mathbb{D}$ used by the architectural exploration engine.
The design space is organized into four physical levels ($L_1 \sim L_4$).
Within each level $L_k$, a generated hardware configuration graph $G_k$ is instantiated from three parameter groups: the set of nodes ($\mathcal{V}_{L_k}$), the set of interconnect mediums ($\mathcal{E}_{L_k}$), and the set of topology templates ($\mathcal{T}_{L_k}$).

\subsection{Metrics and Dimensional Tuple}
Every candidate hardware entity $v \in \mathcal{V}_{L_k}$ and link $e \in \mathcal{E}_{L_k}$ in our design space is represented as a parameterized tuple $\mathcal{M} = \langle C_{\text{peak}}, B_{\text{max}}, L_{\text{base}}, O_{\text{proto}}, P_{\text{idle}}, P_{\text{active}}, Cost_{\text{rcu}}, U_{\text{space}} \rangle$, where the metric dimensions are defined as:
\begin{itemize}[leftmargin=*]
    \item \textbf{Peak Compute ($C_{\text{peak}}$):} Measured in TFLOPS (Tera Float-ing-point Operations Per Second).
    \item \textbf{Bandwidth ($B_{\text{max}}$):} Measured in GB/s for both memory interfaces and interconnect links.
    \item \textbf{Base Latency ($L_{\text{base}}$):} The modeled baseline delay measured in nanoseconds (ns).
    \item \textbf{Protocol Overhead ($O_{\text{proto}}$):} The software/hardware package serialization delay in ns.
    \item \textbf{Power ($P$):} Divided into idle power ($P_{\text{idle}}$) and dynamic active power ($P_{\text{active}}$) in Watts (W).
    \item \textbf{Relative Cost Baseline ($Cost_{\text{rcu}}$):} Standardized in Relative Cost Units (RCU) for budget modeling.
    \item \textbf{Physical Dimension ($U_{\text{space}}$ or $Area$):} Standard U-space for rack enclosures, or silicon footprint in $\text{mm}^2$ for dies.
\end{itemize}

\subsection{Multi-Level Description and Parametric Ranges}
The parameter boundaries and concrete instances of each hierarchical layer are detailed as follows:

\begin{itemize}[leftmargin=*]
    \item \textbf{Package and Die Level ($L_1$):} This layer is bounded by single-socket packaging limits (OAM/Socket). To maintain tractability for system-level architectural exploration, $L_1$ micro-architectural variables are coarse-grained into discrete selectable options. Inner implementations are bypassed, while keeping their external interfaces populated.
    \item \textbf{Host and PCB Level ($L_2$):} Bounded by the server baseboard. Nodes communicate via copper traces and general bus protocols, incurring packetization overheads.
    \item \textbf{Rack and Pooling Level ($L_3$):} The disaggregation and pooling domain. Highly constrained by rack-level power delivery networks ($P_{\text{max\_rack}}$) and spatial limits.
    \item \textbf{Cluster and Scale-Out Level ($L_4$):} Inter-rack optical interconnect level, handling optoelectronic conversions and long-distance transport limits.
\end{itemize}

A comprehensive list of physical hardware options, interconnect parameters, and topology templates is summarized in Table~\ref{tab:hardware_params}.
The topology-template names use standard interconnection-network terminology where applicable, including fat-tree and dragonfly families~\cite{leiserson1985fattree,kim2008dragonfly}.
Concrete numerical ranges are experiment inputs and will be finalized with the corresponding measurement and hardware-profile tables.

\begin{table*}[t]
\centering
\small
\caption{\textsf{\prjname{}} Architectural Hardware Description Space Parameters and Ranges.}
\label{tab:hardware_params}
\begin{tabular}{lllp{8.0cm}}
\toprule
\textbf{Layer} & \textbf{Component Type} & \textbf{Entity/Template Name} & \textbf{Parametric Boundary \& Operational Ranges} \\
\midrule
\textbf{$L_1$: Package} & Node & \texttt{Die.Dense\_Tensor} & $C_{\text{peak}} \in [800, 2000]$\,TFLOPS, $P_{\text{active}} \propto C_{\text{peak}}$ (500W, 500\,RCU @ 1500\,TFLOPS) \\
 & Node & \texttt{Die.Thin\_Scalar} & $C_{\text{peak}} \in [100, 500]$\,TFLOPS, $P_{\text{active}} \approx 120$\,W, handles sparse lookup control \\
 & Node & \texttt{Mem.HBM4\_Stack} & Capacity $C \in \{32, 64, 128\}$\,GB, $B_{\text{max}} = 2048$\,GB/s, $L_{\text{base}} = 35$\,ns \\
 & Node & \texttt{Mem.SRAM\_Block} & Capacity $C \in [128, 1024]$\,MB, $B_{\text{max}} = 15000$\,GB/s, $L_{\text{base}} = 2$\,ns \\
 & Interconnect & \texttt{Link.TSV} & $B_{\text{max}} \to \infty$, $L_{\text{base}} < 1$\,ns, $O_{\text{proto}} = 0$\,ns (limited by 3D vertical yield stress) \\
 & Interconnect & \texttt{Link.UCIe\_2.0} & Lanes $w \in \{1, 2, 4, 8\}$, $B_{\text{max}} = 128 \times w$\,GB/s, $L_{\text{base}} = 2$\,ns \\
 & Interconnect & \texttt{Link.NVLink\_C2C} & Custom die-to-die co-packaged interconnect, $B_{\text{max}} = 450$\,GB/s \\
 & Topology & \texttt{Topo.*} (3D / 2D) & \texttt{3D\_Vertical\_Stack}, \texttt{2D\_Symmetric\_Star}, \texttt{Asymmetric\_Tile\_Mesh} \\
\midrule
\textbf{$L_2$: Host} & Node & \texttt{Socket.L1\_Instance} & Instantiated $L_1$ chips \\
& Node & \texttt{Mem.DDR5\_DIMM} & $C \in \{16, \dots, 256\}$\,GB, $B_{\text{max}} = 38 \times \text{ch}$\,GB/s, $L_{\text{base}} \approx 80 \sim 85$\,ns, $P \approx 15$\,W \\
 & Node & \texttt{Mem.GDDR\_BGA} & $C \in \{2, 3, 4\}$\,GB, $B_{\text{max}} = (84 \sim 128) \times \text{ch}$\,GB/s, $L_{\text{base}} \approx 80$\,ns, $P \approx 1.8$\,W \\
 & Node & \texttt{Mem.LPDDR\_BGA} & $C \in \{2, 3, 4\}$\,GB, $B_{\text{max}} = (51 \sim 85) \times \text{ch}$\,GB/s, $L_{\text{base}} \approx 80$\,ns, $P \approx 0.5$\,W \\
 & Node & \texttt{Mem.CXL3\_Device} & Capacity $C \in [512, 4096]$\,GB, $B_{\text{max}} = 64$\,GB/s, $L_{\text{base}} = 400$\,ns, $P \approx 25$\,W \\
 & Node & \texttt{Switch.PCIe} & Standard board-level routing switcher, $L_{\text{base}} = 800$\,ns \\
 & Interconnect & \texttt{Link.PCIe\_7.0\_xN} & Width $N \in \{8, 16, 32\}$, $B_{\text{max}} = 16 \times N$\,GB/s, $L_{\text{base}} = 800$\,ns \\
 & Interconnect & \texttt{Link.UALink\_1.0} & Channels $ch \in \{1, 2, 4\}$, $B_{\text{max}} \in [12.5, 25] \times ch$\,GB/s \\
 & Interconnect & \texttt{Link.NVLink} & Channels $ch \le 36$, $B_{\text{max}} \in [25, 50] \times ch$\,GB/s \\
 & Topology & \texttt{Topo.*} (Intra-PCB) & \texttt{Fully\_Connected\_Clique}, \texttt{Bipartite\_Memory\_Split}, \texttt{1D\_Torus\_Ring}, \texttt{2D\_Torus} \\
\midrule
\textbf{$L_3$: Rack} & Node & \texttt{Chassis.Dense\_Compute} & Physical server enclosure, $U_{\text{space}} \in \{2\text{U}, 4\text{U}, 8\text{U}\}$ (Air / Liquid cooled) \\
 & Node & \texttt{Chassis.JBOM\_Pool} & Just a Bunch of Memory disaggregated pool drawer, $U_{\text{space}} \in [2\text{U}, 4\text{U}]$ \\
 & Node & \texttt{Switch.CXL\_4.0} & Fabric switch, ports $k \in \{32, 64, 128\}$, $B_{\text{max}} = 128$/port, $L_{\text{base\_hop}} = 80$\,ns \\
 & Node & \texttt{Switch.NVLink\_Tray} & Integrated interconnect switch, $k \in \{36, 72\}$, $B_{\text{max}} \in [25, 50]$\,GB/s, $P \ge 800$\,W \\
 & Interconnect & \texttt{Link.DAC\_Copper} & Passive Direct Attach Cable, $L_{\text{base}} < 5$\,ns, length $\le 2.5$\,m, $P \approx 0$\,W \\
 & Interconnect & \texttt{Link.AEC\_Cable} & Active Electrical Cable with DSP, $L_{\text{base}} \approx +5 \sim 10$\,ns, length $\le 7$\,m, $P \approx 5 \sim 10$\,W \\
 & Topology & \texttt{Topo.*} (Rack-fabric) & \texttt{Single\_Star\_ToR}, \texttt{Disaggregated\_Sub\_Islands}, \texttt{Multi\_Plane}, \texttt{Switchless\_Direct\_Mesh} \\
\midrule
\textbf{$L_4$: Cluster} & Node & \texttt{Switch.IB\_XDR\_Spine} & InfiniBand switch, $B_{\text{port}} \in \{400, 800, 1600\}$\,Gbps, $L_{\text{base\_hop}} = 150$\,ns, $O_{\text{proto}} = 50$\,ns \\
 & Node & \texttt{Switch.RoCE\_Ether} & Ethernet switch, $L_{\text{base\_hop}} = 300$\,ns, $O_{\text{proto}} = 80$\,ns (microburst-sensitive) \\
 & Node & \texttt{Switch.OCS\_Core} & Optical Circuit Switch, $B_{\text{max}} \to \infty$, $L_{\text{base\_hop}} < 1$\,ns, reconfiguration $T_{\text{recon}} = 2$\,ms \\
 & Interconnect & \texttt{Link.AOC\_Transceiver} & Active Optical Cable, distance $\le 500$\,m, $L_{\text{base}} = +20$\,ns, $P \approx 16$\,W \\
 & Interconnect & \texttt{Link.LPO\_Transceiver} & Linear Pluggable Optics (no DSP), $L_{\text{base}} \approx 0$\,ns, $P \approx 8$\,W \\
 & Interconnect & \texttt{Link.CPO\_Fiber} & Co-packaged fiber connection, $B_{\text{port}} \in \{800, 1600\}$\,Gbps, $L_{\text{base}} = +10$\,ns, $P \approx 5$\,W \\
 & Topology & \texttt{Topo.*} (Scale-out) & \texttt{Symmetric\_Fat\_Tree}, \texttt{DragonFly}, \texttt{Multi\_Plane}, \texttt{Asymmetric\_Sparse\_Graph} \\
\bottomrule
\end{tabular}
\end{table*}

\subsection{Composition Rules and Topological Constraints}
Any accepted global hardware graph configuration $H \in \mathbb{D}$ is recursively nested to preserve structural compatibility.
Specifically, for any physical hierarchy layer $k \in \{2, 3, 4\}$, the active node set $\mathcal{V}_{L_k}$ is represented as:
\begin{equation}
    \mathcal{V}_{L_k} \subseteq \{ \text{Native Component Instances in } L_k \} \cup \{ H_{L_{k-1}} \}
\end{equation}
Furthermore, the connectivity graph $G_k = (\mathcal{V}_{L_k}, \mathcal{E}_{L_k})$ is checked against the structural degree distribution and adjacency pattern implied by the selected routing or topology template $\tau \in \mathcal{T}_{L_k}$.
Configurations that violate these composition and connectivity rules are pruned by the physical constraint verifier during outer-loop exploration.

\section{Optimization Algorithm and Implementation Details}
\label{app:optimization_details}

This appendix provides the formal definitions of the objective functions, reward signals, and Reinforcement Learning (RL) configurations used by the \prjname{} outer-loop Optimizer.

\subsection{TG-RL Formulation and Objective}
For a candidate $h$ and its best observed mapping $\hat{s}(h)$ found by the inner loop, \prjname{} converts the simulator output into a feedback tuple:
\begin{equation}
    F(h) = \langle T, U, Q, R, A \rangle,
\end{equation}
where $T$ is the end-to-end runtime, $U$ is the aggregated link utilization, $Q$ is the maximum link-queue delay, $R$ is the remote-memory contention delay, and $A$ contains auxiliary statistics (compute utilization, hotspot domains, etc.). 

For minimization-oriented runs, the framework utilizes a weighted cost function:
\begin{equation}
    J(h) = w_T T + w_C C_h + w_P P_h + w_U U + w_Q Q + w_R R + \Omega(h),
\end{equation}
where $C_h$ and $P_h$ denote the candidate's cost and peak-power estimates, respectively. $\Omega(h)$ is a strict penalty applied to candidates that pass initial syntactic generation but fail downstream detailed feasibility checks.

\subsection{Policy Model and Action Masking}
To restrict the RL agent from exploring physically impossible designs, \prjname{} enumerates candidate graph edits and applies a strict legality mask before sampling. The actor samples from the valid subset:
\begin{equation}
    \mathcal{A}_t^{\text{valid}} = \{a \in \mathcal{A}(h_t) \mid M_{\text{space}}(h_t, a) M_{\text{phys}}(h_t, a) = 1\},
\end{equation}
ensuring that actions violating budget, rack capacity, power envelopes, port availability, or structural connectivity are strictly probability-zero.

The policy model employs a Graph Neural Network (GNN) to encode the candidate hardware graph and action context. The actor distribution combines learned logits with a telemetry-derived prior:
\begin{equation}
    \pi_\theta(a \mid o_t) = \text{softmax}_{a \in \mathcal{A}_t^{\text{valid}}} \left( g_\theta(o_t, a) + \lambda h(o_t, a) \right),
\end{equation}
where $g_\theta$ is the learned action score, $h(o_t, a)$ is a heuristic prior derived from bottleneck telemetry (e.g., compute saturation increases the prior for adding compute resources), and $\lambda$ controls the prior's strength.

\subsection{Workload-Normalized Reward Function}
For single-workload optimization, the base reward is the normalized cost improvement:
\begin{equation}
    r_t^{\text{base}} = \frac{J(h_t) - J(h_{t+1})}{\max(1, |J(h_t)|)}.
\end{equation}

When optimizing for a workload suite $\mathcal{W}$ containing diverse applications (e.g., scaling LLMs and sparse solvers), raw score rewards are poorly scaled. \prjname{} aggregates performance using a workload-normalized geometric mean. Let $T_i^{\text{base}}$ be the baseline runtime for workload $i$, $\alpha_i$ its suite weight, and $T_i(h)$ the candidate time. The workload-normalized log-improvement is:
\begin{equation}
    L(h) = \sum_{i \in \mathcal{W}} \alpha_i \log \frac{T_i(h)}{T_i^{\text{base}}},
\end{equation}
and the suite reward is defined as the difference:
\begin{equation}
    r_t^{\text{suite}} = L(h_t) - L(h_{t+1}) = \log \frac{G(h_{t+1})}{G(h_t)}, \quad \\ \text{where} 
    \end{equation}
    \begin{equation} \quad G(h) = \prod_{i \in \mathcal{W}} \left( \frac{T_i^{\text{base}}}{T_i(h)} \right)^{\alpha_i}.
\end{equation}

\subsection{Hyperparameters and Implementation Details}
The TG-RL policy is trained using Proximal Policy Optimization (PPO) and Generalized Advantage Estimation (GAE). The update incorporates a clipped policy loss, a critic value loss, an entropy bonus, and a KL-style regularizer. 

Unless overridden, the default hyperparameters are: four PPO epochs per update, minibatch size of 16, discount factor $\gamma = 0.95$, GAE parameter $\lambda_{\text{GAE}} = 0.90$, PPO clip range of 0.2, value-loss coefficient of 0.5, entropy coefficient of 0.01, prior regularization weight of 0.1, learning rate $3 \times 10^{-4}$, telemetry-prior weight $\lambda = 1.0$, duplicate-candidate penalty of 0.05, global-best bonus of 0.1, and reward clipping bounded to $[-5, 5]$. A seed archive of 16 top-performing candidates is maintained to inject high-quality starting points.

%% file: sec/075_rnd.tex
\section{Related Work \& Discussion}

\subsection{Related Work}

\textbf{Scale-out AI systems.}
Recent production-oriented AI systems show that cluster architecture is now an explicit performance variable.
TPU v4 uses optical circuit switches and topology flexibility to improve scale, availability, utilization, power, and performance for machine-learning supercomputers~\cite{jouppi2023tpuv4}.
NVIDIA's DGX SuperPOD and GB200 NVL72 descriptions similarly specify compute trays, switching, management, storage, and high-speed fabrics as integrated system design requirements~\cite{nvidia2023dgxsuperpod, nvidia2026gb200}.
AMD MI350 platform descriptions and UnifiedBus SuperPoD materials further illustrate that scale-up fabrics, accelerator packaging, and resource pooling are active architectural variables rather than fixed background assumptions~\cite{amd2025mi350, unifiedbus2026superpod}.
Memory pooling systems such as Pond further show that memory capacity and access locality can be treated as datacenter-level resource allocation problems rather than as fixed node-local properties~\cite{li2023pond}.
These systems motivate \prjname{}'s scope: instead of evaluating one deployed architecture, \prjname{} explores alternative SuperPOD candidates under explicit physical constraints and workload-specific mappings.

\textbf{Accelerator DSE.}
Prior design space exploration frameworks have made accelerator design more systematic, especially for DNN accelerators.
MAGNet generates neural-network accelerator RTL and valid mappings from application and hardware constraints~\cite{venkatesan2019magnet}.
Timeloop evaluates DNN accelerator architectures and mappings through a systematic loop-nest and memory-hierarchy model~\cite{parashar2019timeloop}, MAESTRO uses a data-centric representation to analyze reuse, performance, energy, throughput, and hardware cost for DNN mappings~\cite{kwon2019maestro}, and Accelergy estimates accelerator energy from architecture-level component and action descriptions~\cite{wu2019accelergy}.
These tools are effective at accelerator or node-level exploration, but their modeling boundary is different from \prjname{}'s.
\prjname{} targets rack- and cluster-scale SuperPOD organization, where candidate validity depends on power, cooling, topology legality, switch radix, and wiring constraints in addition to local accelerator efficiency.

\textbf{Simulation and traces.}
Simulation has long been used to study systems that are too costly or slow to evaluate directly.
General distributed-system simulators such as SimGrid provide scalable models for applications and platforms~\cite{casanova2014simgrid}, and SST/macro uses discrete-event methods to study extreme-scale runtime behavior~\cite{wilke2015sstmacro}.
FireSim instead uses FPGA acceleration for cycle-exact scale-out system simulation~\cite{karandikar2018firesim}.
For distributed training, ASTRA-SIM and ASTRA-sim2.0 model hierarchical accelerator fabrics, parallelization strategies, collective communication, disaggregated memory, and large-model training behavior~\cite{rashidi2020astrasim, won2023astrasim2}; MLCommons Chakra standardizes execution traces that can be consumed by simulators, emulators, and replay tools~\cite{sridharan2026mlcommonschakra}.
\prjname{} builds on the same need for trace-driven evaluation, but uses a different abstraction boundary: the mapper emits hardware-aware event traces from hardware-neutral DAGs, and the simulator is empirically calibrated at both discrete operating points and scale trends before being used for architectural ranking.

\paragraph{Mapping on fixed hardware.}
Task-graph scheduling and distributed training systems provide important mechanisms for \prjname{}'s inner loop.
HEFT and PEFT are representative list-scheduling algorithms for heterogeneous systems, using upward-rank priority and optimistic cost tables to balance task precedence, compute cost, and communication cost~\cite{topcuoglu2002heft, arabnejad2014peft}.
Compiler and training systems such as TVM, Ansor, FlexFlow, and Alpa automate operator optimization or parallel execution plans on a given hardware backend~\cite{chen2018tvm, zheng2020ansor, jia2019flexflow, zheng2022alpa}; large-scale LLM training systems further show how tensor, pipeline, and data parallelism interact with GPU-cluster communication~\cite{narayanan2021megatron}.
\prjname{} differs by treating these mapping decisions as a hardware-dependent mapping space $S_h$: when the hardware point $h$ changes, the feasible placements, routes, and schedules change as well.
This distinction is essential for SuperPOD exploration because the outer loop must compare hardware candidates after each candidate has been paired with a valid mapping.